# Parallelization of Network Dynamics Computations in Heterogeneous Distributed Environment

Oleksandr Sudakov and Volodymyr Maistrenko

*Abstract*— **This paper addresses the problem of parallelizing computations to study nonlinear dynamics in large networks of non-locally coupled oscillators using heterogeneous computing resources. The proposed approach can be applied to a variety of nonlinear dynamics models with runtime specification of parameters and network topologies. Parallelizing the solution of equations for different network elements is performed transparently and, in contrast to available tools, does not require parallel programming from end-users. The runtime scheduler takes into account the performance of computing and communication resources to reduce downtime and to achieve a quasi-optimal parallelizing speed-up. The proposed approach was implemented, and its efficiency is proven by numerous applications for simulating large dynamical networks with $10^3$-$10^8$ elements described by Hodgkin–Huxley, FitzHugh–Nagumo, and Kuramoto models, for investigating pathological synchronization during Parkinson's disease, analyzing multi-stability, for studying chimera and solitary states in 3D networks, etc. All the above computations may be performed using symmetrical multiprocessors, graphic processing units, and a network of workstations within the same run and it was demonstrated that near-linear speed-up can be achieved for large networks. The proposed approach is promising for extension to new hardware like edge-computing devices.**

## I. INTRODUCTION

INVESTIGATION of dynamics in large networks of nonlinear and non-locally coupled oscillators is a rapidly growing field of mathematics that finds its application in various branches of engineering, physics, chemistry, biology, sociology, economics, etc. One of its prevalent applications is related to neuroscience, such as the simulation of biological neuronal networks [1], and novel applications like cyber-physical-vehicle systems [2] are emerging. Neuromorphic computing [3] includes novel information processing techniques inspired by the functions of the human brain. Computer simulation of processes in neuronal and other networks can potentially provide information about networks' behavior that is difficult to obtain through other investigation techniques. This information includes oscillators' synchronization at micro-, meso- or macro-levels, simultaneous investigation of different processes, taking into account very large populations of oscillators, etc. Investigation of dynamics in non-locally coupled networks of nonlinear oscillators has revealed phenomena called chimera and solitary states. Many authors consider such states to be important in various applications such as the functions of the human brain [4], etc. Computer simulations of large neuronal or other networks require a large number of neurons, compartments or other oscillators to be taken into account (usually $10^3$-$10^9$), resulting in huge computing resource requirements. Advances in parallel hardware and software allow simulation of relatively large systems [5] but simulation of complicated models and large non-locally coupled networks remains a challenge especially in heterogeneous systems that include different hardware.

Available software packages for simulating large dynamical networks, such as *NEURON* [6], *GENESIS* [7], *BRIAN* [8], *NEST* [9], *PSICS* [10], and others, are primarily designed for modeling biological neuronal networks. Detailed reviews of the most popular packages are available [11]. These packages extensively support compartmental and spiking networks, while support for large coupled oscillator networks is very limited. A compartmental network is described by a set of explicitly specified differential equations, coupled through certain dynamical variables. Compartmental networks are usable for up to tens of oscillators. In a spiking network, the differential equations are specified for a limited set of oscillator classes, e.g. neurons, compartments, synapses. A large number of such oscillator instances are interconnected via synaptic links but have no access to the dynamical variables of each other. Instead, spiking events are sent to postsynaptic oscillators, triggering them when certain dynamical variables cross thresholds in presynaptic oscillators. In a large number of network models (including those in neuroscience, see Section 2) oscillators are continuously coupled through certain dynamical variables and these networks are not spiking ones. The only neuronal network simulator we found that supports large coupled oscillator networks by regular means is *BRIAN2*.

These packages introduce specialized languages for model definitions (*NMODL* for *NEURON*, *NESTML* for *NEST*, etc.) that are usually translated into C/C++ and compiled into

This work was supported in part by the National Academy of Sciences of Ukraine

*Corresponding author: Oleksandr Sudakov*

Oleksandr Sudakov was with Taras Shevchenko National University of Kyiv, Volodymyrska Str. 64/13, Kyiv, Ukraine, 01601. He is now with the Technical Center, National Academy of Sciences of Ukraine, Pokrovs'ka Str., 13, 04070 Kyiv, Ukraine (e-mail: saa@knu.ua).

Volodymyr Maistrenko is with the Technical Center, National Academy of Sciences of Ukraine, Pokrovs'ka Str., 13, 04070 Kyiv, Ukraine (e-mail: maistren@nas.gov.ua).

standalone programs or shared libraries. Models are run using high-level programming languages like Python or HOC to specify parameters, link topologies, visualization, and analysis of results. Network topologies may be specified via equations (*BRIAN2*) or coded in a high-level language by manually creating oscillators and specifying their interconnections. Such approaches perform well for small networks, but for networks with more than approximately $10^6$ oscillators, they frequently result in huge C/C++ source files (hundreds of megabytes), extremely long compiling times (days), long preparation runs (tens of hours) and huge memory consumption (hundreds of gigabytes). Long execution times are particularly detrimental for bifurcation analysis when the results are used as initial conditions for subsequent runs with different parameters and topologies. In such cases, users have to examine or edit the generated C/C++ code and tweak high-level language codes to achieve successful runs and reasonable simulation times.

Simulation of large networks in a reasonable time requires parallel processing. *BRIAN2* supports automatic parallelization via the OpenMP backend and graphics processing units (GPU) backends. Other packages require manual distribution of oscillators across processing elements and support different hardware. *NEURON* (*CoreNEURON* extension) supports symmetric multiprocessing, a network of workstations via MPI, and multiple GPUs within the same run. *NEST* supports MPI and multiprocessing. Most backend implementations do not support adaptive integration steps for differential equation solvers, lack support for heterogeneous resources, or inefficiently utilize them. They are only efficient for spiking networks of integrate-and-fire oscillators when analytical solutions of differential equations are available. Parallel simulation of large coupled oscillator networks is inefficient, if even possible, with such tools. That is why all cases found in the literature describe the application of the tools mentioned above for networks with a maximum of approximately $10^5$-$10^6$ oscillators, and investigators of large coupled oscillator networks use their own simple serial C/C++/Fortran codes. Thus, a new approach for fast computations of large coupled oscillator networks on various hardware requiring minimum programming efforts is needed.

In this paper, we present an approach for transparently parallelizing computations of large nonlinear dynamical networks with non-local coupling. This approach was implemented for massive simulations of large nonlinear dynamical networks with arbitrary link topologies. It was successfully applied for the fast simulation of such networks described by Terman-Rubin, Kuramoto-Sakaguchi and FitzHugh-Nagumo models with $10^3$-$10^8$ elements on various types of hardware, including networks of workstations and computing clusters with message passing interface (MPI), multiprocessor shared memory systems with OpenMP, and graphics processing units (GPUs). Different hardware facilities can be used simultaneously within a single run to simulate a single network. On heterogeneous hardware, computations are scheduled automatically to achieve better performance. The proposed approach may be useful not only for commodity hardware but also for high-performance computing clusters, grids, and cloud systems. It can also be extended to modern high-performance edge devices. The existing tools do not have all of these features, as proven by performance comparison with *BRIAN2*.

The paper is organized as follows: Section 2 briefly describes different types of dynamical network models. Section 3 describes the algorithms and data structures of the proposed approach. Section 4 discusses selected applications of the proposed approach. Section 5 is devoted to performance analysis, and Section 6 provides discussions and conclusions.

## II. Dynamical Networks

There are a variety of dynamical network models available [12], many of which are related to neuroscience. Microscopic neuronal network models describe the specific characteristics of cells or cell compartments and consider membrane potentials, ion channel currents, ion concentrations, etc. Mesoscopic models [13] describe characteristics of brain tissues that depend on the collective behavior of cell groups, such as local field potentials (LFP), electroencephalography (EEG), and magnetoencephalography (MEG) signals, etc. Macroscopic models [14] aim to describe the functions of the entire brain or large brain areas such as cognition, memory, and neuronal pathologies. Dynamical models explain the behavior of networks through differential or recurrent equations. Stochastic models represent network behavior as random variables or processes. The most commonly utilized descriptions involve systems of nonlinear differential equations that incorporate random variables or processes.

Realistic models describe actual physical parameters of cells or compartments, such as membrane voltage, ion channel currents and conductivities, ion and other substance concentrations, etc. These neuronal models are mostly variations of the Hodgkin-Huxley model [15], as shown below:

$$
\begin{aligned}
&C_i \frac{dU_i}{dt} = \sum_j^M I_j(U_i, g_j, n_j) + I_e\left(U_i, \sum_k^N L_{ik} s_{ik}\right) + I_{app},\\
&\frac{dg_j}{dt} = f_j(U_i, g_j, n_j),\\
&\frac{dn_j}{dt} = h_j(U_i, g_j, n_j),\\
&\frac{ds_{ik}}{dt} = p_{ik}(U_i, s_{ik}, n_j),
\end{aligned}
\qquad (1)
$$

where $\frac{d}{dt}$ is the time derivative; $i$ and $k$ enumerate network elements (neurons or compartments) $[0 \dots N)$; $j$ enumerates the ion channels in a neuron or compartment $[0 \dots M)$; $k$ enumerates external neurons or compartments coupled with the $i$-th one; $C_i$ is the membrane capacitance; $U_i$ is the membrane voltage; $g_j$ is the conductivity of ion channels; $n_j$ is the concentration of ions; $I_j$ describes the ion currents that are nonlinear functions of membrane voltage, ion channel conductivities, and ion concentrations; $I_e$ describes the external currents induced in the membrane by other neurons or compartments (synaptic transmission, etc.); $s_{ik}$ is the synaptic conductivity from the $k$-th to the $i$-th neuron or compartment; $L_{ik}$ is the coupling matrix with elements equal to 1 or 0 depending on whether the $k$-th network element affects the $i$-th network element or not; $f_j, h_j, p_{ik}$ describe changes of conductivities and concentrations over time and are related to ion-channel and synapse activation, inhibition, plasticity, etc.; $I_{app}$ is the applied current due to random noise or current needed to be applied for the correct solution of differential equations. Dynamical variables (e.g., membrane voltages, channels' conductivities) change over time depending on the initial values of these variables and equation parameters.

Solutions of realistic models show good correspondence with experiments. However, realistic models typically involve a relatively large number of differential equations per oscillator (usually 3-12), a significant number of parameters, and nonlinear functions in the right-hand parts of the differential equations, making it difficult to solve and investigate such models. Under certain conditions, these differential equations may become stiff, meaning the solution changes rapidly, and conventional numerical integration methods may fail to obtain an accurate solution.

To simplify the investigation of realistic network models, many phenomenological models have been introduced. Such models describe not the behavior of real physical systems but rather some observable properties of such systems or other systems that to some extent behave similarly to the systems under investigation, e.g., generate spiking, bursting, or other oscillations similar to neurons. Such models have few (typically two) dynamical variables that describe slow and fast dynamics.

An example of a phenomenological model is the FitzHugh–Nagumo model [16], which is as follows:

$$\varepsilon \frac{dx_i}{dt} = x_i - \frac{x_i^3}{3} - y_i + \sum_j^N L_{ij} s_{ij}\big(x_j(t-\tau) - x_i\big) + I_{app}$$
$$\frac{dy_i}{dt} = x_i + a + Dn(t), \qquad (2)$$

where $\frac{d}{dt}$ is the time derivative; $i$ and $j$ enumerate the network elements $[0 \dots N)$; $x$ is the fast variable; $y$ is the slow variable; $s_{ij}$ is the coupling strength; $L_{ij}$ is the coupling matrix; $\tau$ is the delay in time; $n(t)$ is the random noise; $\varepsilon, a$, and $D$ are the model parameters; $I_{app}$ is the applied current. This model is derived from the Van der Pol model of a tube generator. Such a model is widely used for the investigation of synchronization.

Phase models are special types of phenomenological models that describe not the behavior of physical characteristics, but the behavior of phase in a phase space of some system. Kuramoto-type models [17] are commonly known models of phase oscillators. The Kuramoto-Sakaguchi model with inertia is as follows:

$$m \frac{d^2 \varphi_i}{dt^2} + \varepsilon \frac{d\varphi_i}{dt} = \mu \sum_j^N L_{ij} sin\big(\varphi_j - \varphi_i - \alpha\big), \qquad (3)$$

where $\frac{d^2}{dt^2}$ and $\frac{d}{dt}$ are the second and first time derivatives; $i$ and $j$ enumerate network elements $[0 \dots N)$; $\varphi_i$ is the phase of the $i$-th oscillator; $m$ is the mass parameter; $\varepsilon$ is the damping parameter; $\mu$ is the coupling strength; $L_{ij}$ is the coupling matrix, and $\alpha$ is the phase lag.

Each of the described models has a certain number of dynamical variables, a certain number of parameters, and a coupling matrix. Equations for different network elements are coupled via some dynamical variables. The number of differential equations scales proportionally to the number of network elements $N$. The number of operations scales from $O(N)$ to $O(N^2)$ depending on the coupling topology. The coupling matrix for a network with $N$ elements scales as $N^2$ and may become too large to fit into memory.

Integration of differential equations, similar to equations (1)-(3), requires the runtime specification of initial values for dynamical variables, model parameters, and the coupling topology of the network. The integrator outputs a dynamical trajectory, i.e., values of dynamical variables at specified time moments. The trajectory should be further analyzed to generate reports, plots, animations, aggregated results, etc. Network model investigation requires computing thousands of dynamical trajectories. Such massive computations require batch mode operation, and computational software should be controlled with configuration files [18]. One of the best ways to represent the output results is through the generation of aggregated reports accessible via the web or other interfaces [18-21].

## III. PARALLELIZATION

### A. *Parallel Integrator*

The parallel integrator diagram is displayed in Fig. 1a. Each oscillator in the simulated network belongs to one group. The mapping of oscillators to groups is specified via a configuration file. The numbers of oscillators and groups are 64-bit integers that can satisfy a wide range of practical needs. Oscillators in the same group share the same model, i.e., the differential equations, and have the same parameters. Oscillators in different groups may have the same or different parameters and the same or different models. All models are compiled into the integrator to achieve the best performance. The parameters of the models are read from the configuration file using the models' *ReadConfig()* methods.

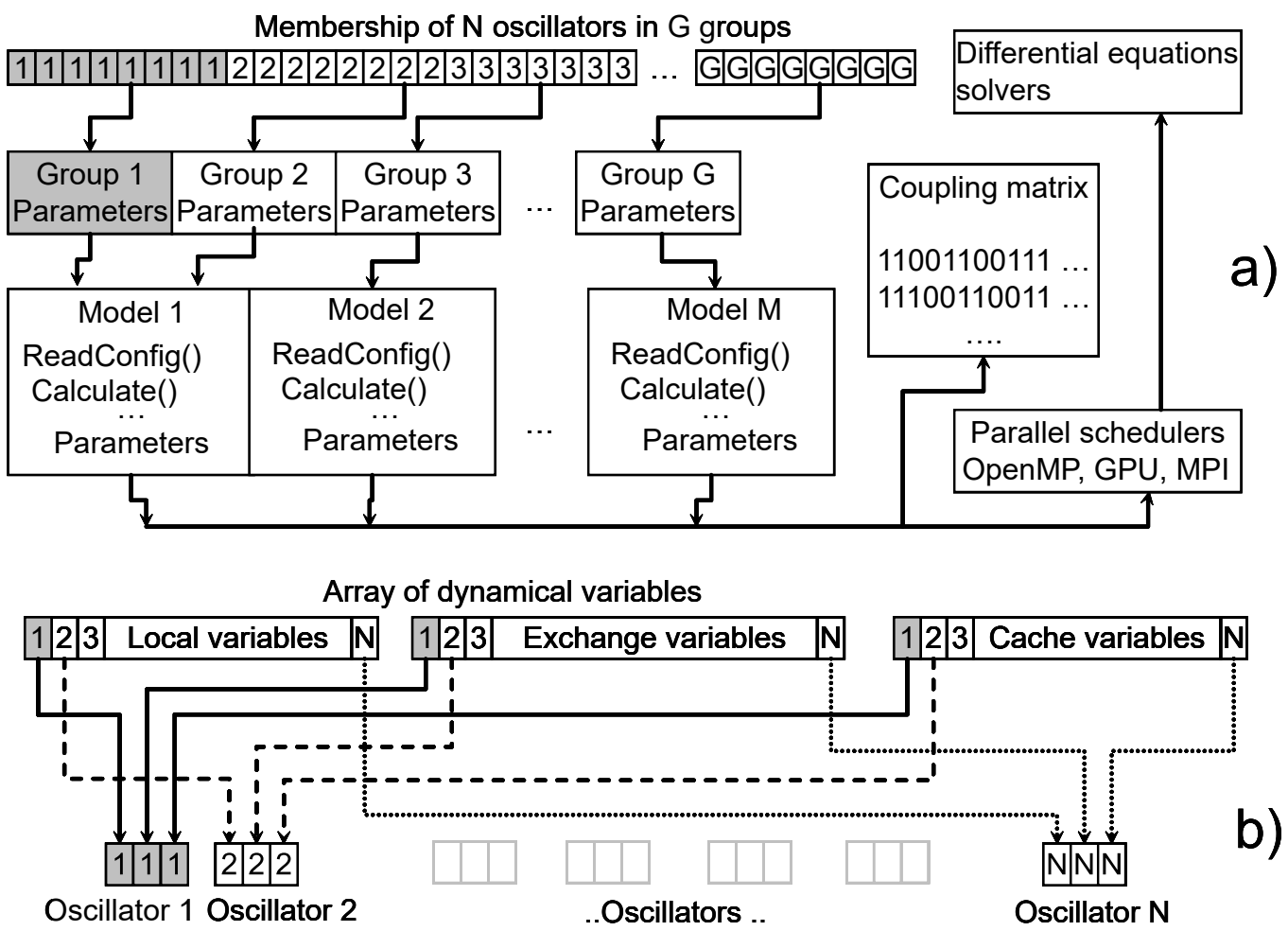

**Fig. 1.** (a) Parallel integrator diagram. (b) Memory layout.

Differential equations for all oscillators can be solved using several differential equation solvers: a Dormand-Prince solver of 3, 5, or 8 order [22] for non-stiff problems with delays, a solver for moderately stiff problems using back differentiation formula [23], a solver for stiff problems using 5-order implicit Runge-Kutta method [24], and a solver with very high precision using Richardson extrapolation formula [22]. The last two solvers are much slower than the others and are recommended only if Dormand-Prince 5-th order or back differentiation solvers fail. A Lyapunov exponents estimator [25] is included as a special solver type. Differential equation solvers call the models' functions *Calculate()* in parallel to compute the right-hand parts of the differential equations for each oscillator, as described below in Subsection D. Dormand-Prince solvers internally support differential equations with delays, maintaining data from previous time steps and performing interpolation. It is always possible to implement delays in models by storing data from previous steps in additional dynamical variables.

Different models can be used in the same dynamical network but these models should be "aware" of which dynamical variables are used for oscillator coupling in different models. The coupling matrix specifies which oscillators are coupled and which are not, and is used by the oscillators' models to determine which oscillators are coupled to the current oscillator. The coupling matrix may be specified at runtime.

### *B. Data Structures*

Models used for computations are compiled into the integrator for performance reasons and implemented in C++. A simple application programming interface (API) is provided for including new models with minimal effort. *Algorithm 1* describes an example of a fully functional Kuramoto-Sakaguchi model (3) implementation with $m = 0$, $\varepsilon = 1$, and $\mu = 1/\Omega$, where $\Omega$ is the count of oscillators coupled to the current one.

The optional structure *KurOscil* describes an oscillator with a single dynamical variable *phi* that is initialized from a provided memory location in the constructor. The structure *KurModel* inherits the recursive template *BaseModel* and defines an oscillator having 0 local variables and 1 exchange variable *phi*. Such oscillators are described by the template *NeuronModelBase<0,1>* structure originally designed for neuroscience purposes and may be used for any models. The structure *KuramotoConfig* describes the model's parameter $\alpha$ that may be specified via the configuration file. The model has two mandatory functions *ReadConfig()* and *Calculate()*. Functions for data input, data output, and others are optional, and defaults are provided for them. The function *ReadConfig()* reads the model's parameter $a$ from the configuration file using the macro *CFG_PARAM()*.

The function *Calculate()* computes the right-hand part of differential equation (3) for the oscillator's data provided in the function's parameter *in*. The variable *cur* represents the current oscillator, which is initialized from a memory location in the array of exchange variables *in.exchange*, provided in the function's argument. The variable *der* represents the derivative of the oscillator's dynamical variable *phi* computed by the *Calculate()* function. This computed value is saved to the memory location *out.exchange* provided in the function's argument. The loop *for* iterates over all oscillators coupled to the current one according to the coupling matrix *in.links*. The iterator *l* returns the numbers of oscillators coupled to the current oscillator. The function *get_exchange_offset* returns the offset in the array of exchange variables (*c_exc*) and a group (*c_grp*) for a coupled oscillator. The variable *c* is initialized from the memory location of the coupled oscillator's dynamical variable and is used for derivatives (3) computations. The keywords __host__ and __device__ are added to the functions that should be called on processors (CPUs) and GPUs, i.e., *Calculate()* and the *KurOscil()* constructor. The described model should be included in a software tree, recompiled, and can be used for computations. In general, any number of dynamical variables and any number of configuration parameters may be specified. Although simple and used in production by the authors, the described model specification API should not be considered ready for public use like APIs [6-12] because it is subject to changes, requires manual recompilation, and development in C++ with knowledge of the framework internals.

**Algorithm 1** Kuramoto model implementation.

```
struct KurOscil {
    double& phi;
    __host__ __device__ KurOscil(double* x):phi(*x){}
};
struct KurModel: public
        BaseModel <KurModel, NeuronModelBase<0,1> >{
    struct KuramotoConfig {double a;} C;
    int ReadConfig(read_config& cfg,
                cfg_section& section, bool strict){
        CFG_PARAM(a); return 1;
    };
    __host__ __device__ void Calculate(const double t,
            const NeuronData& in, const NeuronData& out){
        KurOscil cur(in.exchange);
        KurOscil der(out.exchange); der.phi = 0;
        size_t links_amount = 0; auto l = in.links->begin();
        for(; l != in.links->end(); ++l, ++links_amount){
            size_t c_grp, c_exc;
            get_exchange_offset(*l, c_grp, c_exc);
            KurOscil c(in.all_exchange + c_exc);
            der.phi += sin(c.phi – cur.phi - C.a);
        }
        der.phi /= links_amount;
    };
};
```

The data layout of oscillators is described in Fig. 1b. For performance reasons, each oscillator can contain three types of variables: local, exchange, and cache. Local variables are dynamical variables not used for oscillator coupling, exchange variables are dynamical variables used for coupling, and cache variables store computed results for further usage. For example, derivatives of dynamical variables may be cached for further analysis. The local, exchange, and cache variables are stored in continuous memory areas, which significantly improve the performance of hardware caching, memory accesses, and data transfers. To compute the coupled oscillators in parallel, the exchange variables should be more intensively transferred via communication channels compared to local and cache variables.

### *C. Coupling Matrix Compression*

The coupling matrix (link matrix, Fig. 1a) $L_{ij}$ contains information on how oscillators in the network are interconnected. If the $j$-th (presynaptic) oscillator influences the $i$-th (postsynaptic) oscillator, then $L_{ij} = 1$; otherwise, $L_{ij} = 0$. Therefore, the network links are directed from the $j$- th oscillator to the $i$-th one. The network may be either connected or unconnected. Unconnected subnets of the network may be computed within different runs, but having a large number of small unconnected subnets within a single run makes sense for performance optimization. It is important to note that the coupling matrix specifies the presence or absence of coupling between each pair of network elements, and the binary coupling matrix can describe all possible link topologies. Using a binary matrix to specify the coupling topology imposes no restrictions on the models of links. If coupling exists or can potentially exist between certain oscillators, the links between them can be described by any number of differential equations with any number of dynamical variables and parameters.

The link matrix can be specified at runtime and has a size of $N^2$ for a network with $N$ elements. Despite containing only values of 0 or 1, it can become very large for large networks. Without compression it may require memory larger than the computer system can provide. For example, a network with $10^6$ elements has a coupling matrix with $10^{12}$ elements. If 1 bit is used for a single matrix element, then the matrix size is approximately 125 GBytes. Without compression, it cannot fit into the memory of most existing GPUs.

*Algorithm 2* describes a pseudo-code of the differential compression approach proposed for the coupling matrix [26]. Notation "{}"- specifies ordered evaluation, e.g., $\{y, z\} \leftarrow f(\{g(x), h(y)\})$ means $y \leftarrow f(g(x))$ and then $z \leftarrow f(h(y))$. The coupling matrix $\hat{\mathbf{L}}$ is input by rows. Each row $\hat{\boldsymbol{L}}_{i,}$ is a bit vector with bits set to 1 at the positions equal to the numbers of oscillators influencing oscillator $i$. The first row $\hat{\boldsymbol{L}}_{0,}$ is assigned to a *common vector* $V$. The essential part of the compression procedure, *COMPRESS,* computes a circular shift of the row $\hat{\boldsymbol{L}}_{i,}$ by $i$ positions left ($ROL$) and consecutive $XOR$ operation with the common vector $V$ to produce a *difference vector* $d$. The *row* of the *compressed* coupling matrix $\widehat{\boldsymbol{C}}_{i,}$ is a reference to the *difference table* entry $\widehat{\mathbf{D}}_{j,}$ that contains the difference vector $d$. If the difference vector $d$ for the row $\hat{\boldsymbol{L}}_{i,}$ is not present in the difference table $\widehat{\mathbf{D}}$, it is added to the end of the difference table.

**Algorithm 2** Coupling matrix compression.

COMPRESS($\hat{\mathbf{L}}$)
    $V \leftarrow \hat{\boldsymbol{L}}_{0,}$
    **for** $i = 0, \dots, |\hat{\boldsymbol{L}}_{,\cdot}| - 1$
        $\{d, j\} \leftarrow \{\mathrm{XOR}\ (\mathrm{ROL}\ (\hat{\boldsymbol{L}}_{i,}, i), V), \mathrm{FIND}(\widehat{\mathbf{D}}_{j,} == d)\}$
        $\widehat{\boldsymbol{C}}_{i,} \leftarrow \widehat{\mathbf{D}}_{j,}$ **if** $\mathrm{EXIST}(j)$ **else** $\left\{\hat{\mathbf{C}}_{i,}, \widehat{\mathbf{D}}_{|\widehat{\mathbf{D}}_{\cdot,}|,}\right\} \leftarrow \{d, d\}$
    **return** $V, \widehat{\mathbf{D}}, \hat{\mathbf{C}}$

EXTRACT($i, V, \widehat{\boldsymbol{C}}$)
    **return** $\mathrm{ROR}(\ \mathrm{XOR}(V, \hat{\mathbf{C}}_{i,}), i)$

The idea behind the compression is based on the likely assumption that most oscillators in the network have a similar structure of couplings with the neighboring oscillators. This means that the rows of the coupling matrix for each pair of oscillators should have many common set and unset bits after shifting by the difference in the numbers of these oscillators. Applying $XOR$ to such shifted rows produces a bit vector with very few set bits, which is easy to compress using run-length encoding. For regular networks where most oscillators have similar link structures, there should be very few entries in the difference table. For example, a full mesh topology has a common vector with all bits set and a difference table with a single entry having all bits unset. The decompression procedure, *EXTRACT*, for the compressed topology matrix row $\hat{\mathbf{C}}_{i,}$ is performed by $XOR$ operation with the common vector $V$ and consecutive circular shift by $i$ positions to the right, $ROR$.

Compression requires $O(\max(NBK, NL))$ operations in the worst case, where $N$ is the number of network elements, $L$ is the number of links per oscillator, $B$ is the maximum number of set bits in difference table entries, $K$ is the number of difference table entries, $O(NBK)$ is the number of difference table lookup operations, and $O(NL)$ is the number of XOR operations with the common vector. Decompression $XOR$ procedure requires $O(L)$ operations for $L$ oscillators coupled to the current one, i.e., an average of $O(1)$ operations per link.

Coupling matrix operations were implemented using the BitMagic library [28], which supports high-performance compressed bit vectors. Comparing BitMagic vectors for inequality is a relatively fast operation that does not require comparing all bits, only the positions of the set bits. Circular shift is implemented using the addition of bits' positions modulo $N$. The fast $XOR$ is implemented by comparing the positions of set bits in the argument vectors. Taking all of the above into account, the proposed compression algorithm requires not more than $O(N^2)$ operations for coupling matrix compression, which is performed only once. The compressed matrix size in memory is $O(\max(BK, N))$, which is never larger than the uncompressed matrix size. For example, the compressed link matrix for 3D networks with $10^6$ elements, and a coupling radius of 6 (see Section IV) occupies approximately 16 megabytes of memory, compared to approximately 125 gigabytes for the uncompressed link matrix and about 8 gigabytes for the per-row compressed link matrix. The compression ratio can reach 8000. The compression time for this matrix is about tens of seconds on the processors used for testing (see Section 5). The coupling matrix compressed with this approach fits into the memory of a commodity GPU even for networks with $10^7$-$10^8$ elements. A small difference table and a single common vector provide high hardware cache hit rates and high performance. The decompression requires a constant number of operations per link and increases computing time 2-5 folds compared to uncompressed bit arrays. The main shortcoming of the proposed compression approach is the impossibility to change the link matrix after it is compressed, which is usually not a problem for most applications.

For small networks, per-row compressed bit vectors or uncompressed bit vectors are supported, allowing runtime changes to the coupling matrix in shared memory. Uncompressed bit vectors are potentially the fastest approach when the coupling matrix fits into memory, but become very slow in the opposite case, so they are useful only for relatively small networks with about up to $10^5$ elements. Compression of each coupling matrix row separately reduces the matrix size by approximately an order of magnitude while also reducing performance several folds. Dynamic updates of the link matrix in distributed memory systems require the propagation and synchronization of changes to other devices, significantly reducing performance and requiring additional research for optimization. The tools [6-12] use differential equations to describe neuronal plasticity and other dynamic changes in link strengths. Their link topology is always static during simulation. The presented approach exhibits similar behavior and allows quick restart of the computations from any dynamical state with a changed coupling matrix, as shown in the examples of Sections 4-5.

### D. Parallel Scheduling

Computing the right-hand parts of differential equations like (1)-(3) requires $O(NL)$ operations that are performed several times per integration step, where $N$ is the number of network elements, $L$ is the number of links per oscillator. Since this part of the task is the most computationally intensive, it is executed in parallel for different network elements. The right-hand part of the differential equation for each individual oscillator is computed serially as described by *Algorithm 1*. Computations of derivatives and equation solver algorithms require $O(L)$ and $O(N)$ operations, respectively, and can also be parallelized. However, it usually does not lead to a significant speed-up for large networks. Different hardware may be used for these computations: shared memory multiprocessors (OpenMP), graphic processing units (GPU) and a network of workstations with a message passing interface (MPI). These hardware types may be used simultaneously within the same run for a single network and separate binaries for OpenMP, OpenMP+GPU, OpenMP+MPI, and OpenMP+MPI+GPU are compiled.

Parallel schedulers perform dynamic transparent runtime performance optimization on different hardware. The details of these schedulers without checking errors, empty ranges, and other corner cases are described in the following subsections using *Algorithms 3-5* pseudo-code, where the notation "[]" specifies an integer part. *Algorithm 3* is a scheduler for shared memory multiprocessors with several GPUs. *Algorithm 4* is a scheduler for a network of shared memory multiprocessors. Algorithm 5 is a scheduler for a network of shared memory multiprocessors with GPUs.

A general idea of parallel scheduling is shown in Fig. 2. The initial values of dynamical variables, models' data, and the coupling matrix are read from files. The coupling matrix is compressed using *Algorithm 2*, and the oscillator network is built. For *Algorithms 3* and *5*, the coupling matrix and the models' data are propagated across GPUs. During the dynamical trajectory computations, the integrator updates the dynamical variables of all oscillators; the scheduler distributes the oscillators across processors, GPUs (*Algorithms 3, 5*), and network nodes (*Algorithms 4, 5*); the derivatives of dynamical variables are computed in parallel on the processors, GPUs, and network nodes depending on the hardware and scheduler. Processing elements execute the models' *Calculate()* method of oscillators scheduled to them (like *Algorithm 1*). After the completion of parallel computations, the results are transferred back into the hosts' memories from GPUs (*Algorithms 3, 5*) and are exchanged over the network (*Algorithms 4, 5*). Finally, the scheduler estimates the relative performance of different processing elements and reschedules oscillators across these processing elements to minimize hardware downtime. Thus, scheduling *Algorithms 3, 4*, and *5* have some common and some distinguished parts. Parts specific to GPU schedulers are shown with dashed lines, while parts specific to MPI schedulers are shown in dotted lines (Fig. 2). The following subsections describe the details of *Algorithms 3, 4*, and *5* and their common and distinguished parts.

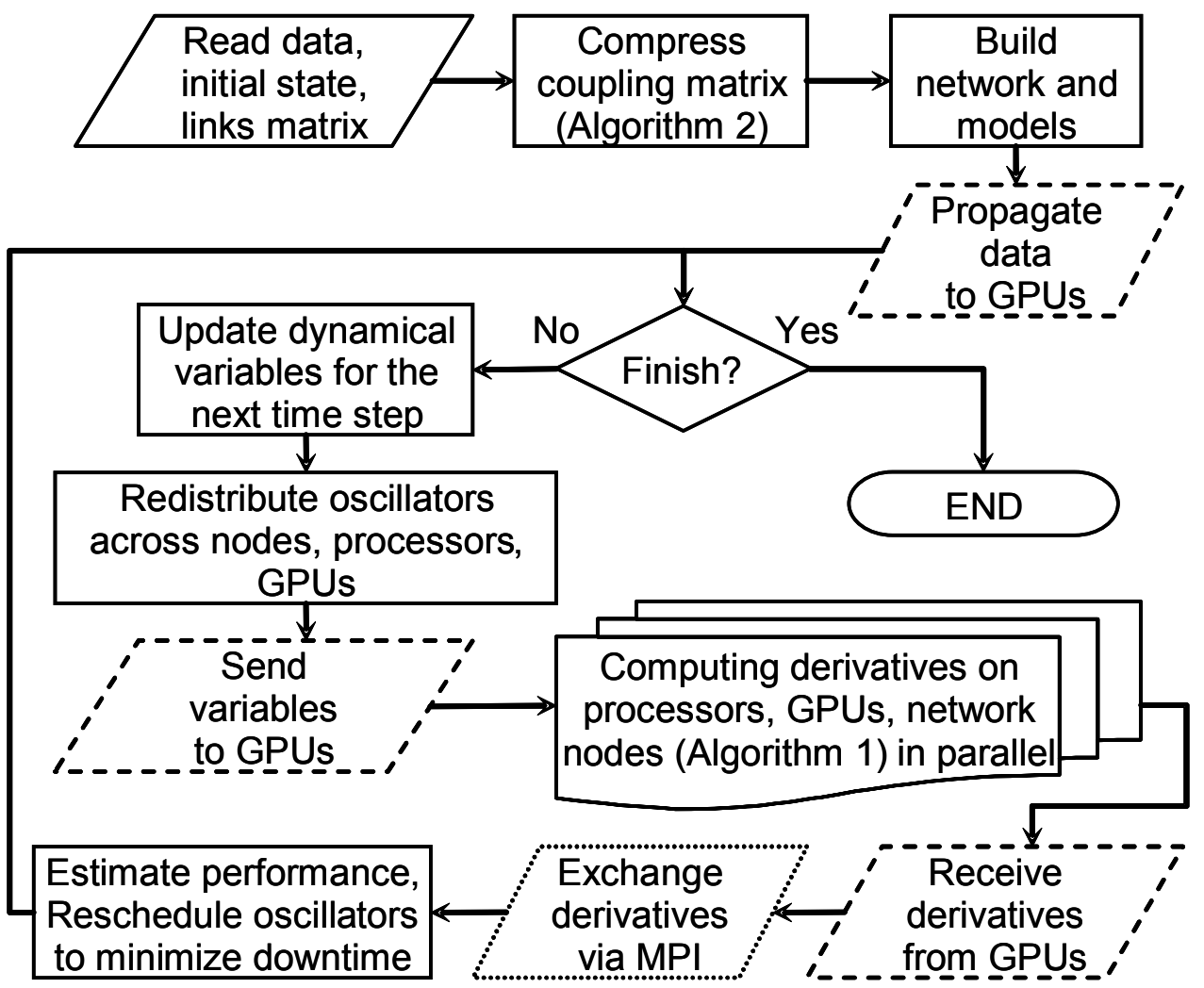

**Fig. 2.** Parallel execution flowchart.

### E. Shared Memory with GPUs

*Algorithm 3* scheduler transparently optimizes runtime performance on a shared memory multiprocessor with several GPUs. The host and GPUs are parts of a distributed system with significantly heterogeneous resources that require data exchange. The scheduler is intended to be used with host processors and GPUs that are compatible in binary data representation and is implemented using the OpenMP standard for shared memory multiprocessing [28] and the CUDA API for NVIDIA GPUs [29].

All initialization data, such as the count of oscillators $N$, mapping of oscillators to groups, models and their parameters $\boldsymbol{M}$, compressed coupling topology matrix $V, \hat{\mathbf{C}}, \hat{\mathbf{D}}$ and others, are read by the host code from configuration files into the host's memory. These data are mapped to the GPUs' memories in the *INIT_G3* procedure using the *CudaHostRegister* API. The *InitKernel* kernels are executed on all GPUs to perform initialization tasks such as creating the models' instances, calling their copy-constructors, updating difference table pointers, etc. Keeping copies of initial data on all parts of the heterogeneous system requires little memory due to the compression approach described above and provides performance optimization. The *INIT_G3* procedure also initializes an even distribution of oscillators with numbers in the range $[B:E)$ across $N_G$ GPUs and the host. The GPU with number $g$ receives oscillators with numbers in the range $[n_g:n_{g+1})$. The host receives oscillators $[n_{N_g}:E)$. The scheduler parameters $\delta_g$ are set to half of the assigned oscillators count.

Parallel execution is performed inside the procedure *OMP_GPU_DRV3* called by the equation solvers. Execution on all available GPUs, *EXEC_GPU*, is started by the host's OpenMP threads in parallel. The remaining parallel threads execute the *Calculate()* method for oscillators assigned to the host. After the host completes its computations, the ready status for all GPUs is saved in the $F_g$ flags. GPUs that complete their jobs before the host have $F_g = 0$, while the rest of the GPUs have $F_g \neq 0$. The call to *cudaDeviceSynchronize()* provides an execution barrier for all GPUs. Finally the numbers of oscillators assigned to the GPUs are increased or decreased by the $\delta_g$ if the corresponding GPU is faster or slower than the host, and numbers of oscillators assigned to the rest of the GPUs and to the host are updated appropriately. GPUs with lower numbers have precedence in scheduling. Convergence is guaranteed by halving the $\delta_g$ values until they reach $\delta_{min}$. The converged state slowly changes, adapting to the varying computing times on different devices.

**Algorithm 3** Parallel scheduler OpenMP+GPU.

INIT_G3($N_G, B, E, \boldsymbol{M}, V, \widehat{\mathbf{C}}, \widehat{\mathbf{D}}$)<br>
  $n_g \leftarrow B + [g(E-B)/(N_G+1)]$ **for** $g = 0, \dots, N_G + 1$<br>
  $\delta_g \leftarrow [(n_{g+1} - n_g)/2]$ **for** $g = 0, \dots, N_G$<br>
  CudaHostRegister($g, \boldsymbol{M}, V, \widehat{\mathbf{C}}$) **for** $g = 0, \dots, N_G - 1$<br>
  InitKernel($g, \boldsymbol{M}, V, \widehat{\mathbf{C}}$) **for** $g = 0, \dots, N_G - 1$<br>
  **return** $n, \delta$

EXEC_GPU($g, B, E, S$)<br>
  cudaMemcpyAsync(exc$_{[B:E)}$, $g$, 0, HostToDevice)<br>
  $e$ ←cudaEventRecord($g$)<br>
  **for** $s = 0, \dots, S-1$<br>
    $\{b, f\} \leftarrow B + [\{s, s+1\}(E-B)/S]$<br>
    cudaMemcpyAsync(loc$_{[b:f)}$, $g, s$,HostToDevice)<br>
    cudaStreamWaitEvent($e$)<br>
    CalculateKernel($[b\colon f), \boldsymbol{M}_{[b:f)}$)<br>
    cudaMemcpyAsync(der$_{[b:f)}$, $g, s$,DeviceToHost)<br>
  **return** der$_{[B,E)}$

OMP_GPU_DRV3($N_G, n, \boldsymbol{M}, S, \delta, \delta_{\min}$)<br>
  **omp parallel for nowait** $g = 0, \dots, N_G - 1$<br>
    der$_{[n_g:n_{g+1})}$ ←EXEC_GPU($g, n_g, n_{g+1}, S$)<br>
  **omp parallel for nowait** $c = n_{N_G} \dots, n_{N_G+1} - 1$<br>
    der$_c$ ←Calculate($\boldsymbol{M}_c$)<br>
  $F_g$ ←cudaErrorNotReady($g$) **for** $g = 0, \dots, N_G - 1$<br>
  cudaDeviceSynchronize($g$) **for** $g = 0, \dots, N_G - 1$<br>
  **for** $g = 0, \dots, N_G$-1<br>
    $n_j \leftarrow n_j + (F_g ? \delta_g : -\delta_g)$ **for** $j = g, \dots, N_G$<br>
    $\delta_g \leftarrow \delta_g/2$ **if** $\delta_g > \delta_{\min}$<br>
  **return** der, $n, \delta$

The scheduler aims to achieve equal computing time on all GPUs and the host to prevent their downtime. Its convergence time is $O(\log N)$, and its operations per iteration are $O(N_G^2)$. The latter is not a problem for a small number of GPUs, which is usually the case. The proposed scheduler does not guarantee optimal performance, but its application has always resulted in shorter execution times compared to static distribution of oscillators across devices.

The procedure *EXEC_GPU* controls execution on the $g$-th GPU using the CUDA streams API. Initially, asynchronous data transfer from the host to the device is started by the *cudaMemcpyAsync()* call for all exchange variables of all oscillators. Then a synchronization event of the exchange variables transfer $e$ is registered by the *cudaEventRecord()* call. Oscillators in the range $[B\colon E)$, assigned to the $g$-th GPU, are evenly distributed across $S$ streams. Each stream asynchronously performs the following actions with oscillators in the range $[b\colon f)$: transfer of local and cache variables ($loc$) from the host to the device; waiting until the exchange variables transfer is completed (*cudaStreamWaitEvent()*); execution of the kernel *CalculateKernel()* on the device for parallel evaluation of the oscillators' *Calculate()* method; transfer of computed derivatives $der$ from the device to the host. This approach enables overlap and pipelining of bidirectional data transfers and computations. Approximately 20 CUDA streams were determined to be optimal for the tested cases.

### *F. Network of Shared Memory Workstations*

*Algorithm 4* scheduler transparently optimizes runtime performance in a network of shared memory symmetric multiprocessors. All available nodes in the network cluster execute the same program. Each instance of the program reads configuration data, constructs a copy of the dynamical network and carries out calculations for the assigned oscillators. The *INIT_N4* procedure initializes an even distribution of $N$ oscillators across $N_N$ nodes. Procedure *OMP_MPI_DRV4* performs parallel computing on the $r$-th node. Oscillators with numbers in the range $[m_r\colon m_{r+1})$ assigned to the $r$-th node are evenly distributed across $N_P$ partitions. Parallel OpenMP threads execute the *Calculate()* method for oscillators of the $p$-th partition with numbers in the range $[b_p\colon e_p)$. When these calculations are completed the exchange of computed derivatives $der$ for the $p$-th partition is initiated by the collective non-blocking MPI-3 call *MPI_Iallgatherv()* [30].

After completing calculations for all oscillators of the $r$-th node, the scheduling parameters are updated using the *MPI_RESCHED4* procedure. Initially, the computation time $t_r$ for the $r$-th node is transmitted to other nodes using the *MPI_Iallgather()* function. The *MPI_Waitall()* function is then called to synchronize data exchange. Computation and data exchange for partitions are overlapped and pipelined. The computation time $t_r$ at the $r$-th node is exponentially averaged over successive runs with a relaxation parameter $\tau$~5. The parameter $\tau$ regulates the stability of rescheduling and the response to changes in computing speed. Large values of the parameter $\tau$ lead to smoother and slower rescheduling. The average computing speed $v$ is calculated by summing the inverse of the average computing times $1/T_r$ for all nodes. The range of oscillators on the $r$-th node is adjusted by the value $(1/T_r - v)N/vN_N$. This value is the average number of oscillators per node scaled by the relative difference of the node's speed from the average speed value. The range adjustment approaches zero as nodes' speed approaches the average speed. Scheduling requires $O(N_N)$ operations. This scheduler consistently delivers better performance compared to the static distribution of oscillators across nodes. The optimal number of partitions corresponds to the equality of partition's computing and exchange times, which significantly depend on the network size, models, computing and communication speed. In practice several hundred partitions provided the best performance for large networks. Non-blocking collective communication provided better performance compared to one-sided communication.

**Algorithm 4** Parallel scheduler OpenMP+MPI.

INIT_N4($N_N, N$)
$m_r \leftarrow [rN/N_N]$ **for** $r = 0, \dots, N_N$
**return** $m$

MPI_RESCHED4($N_N, N, m, t, \tau$)
MPI_Iallgather(MPI_Wtime()$-t_r$)
MPI_Waitall()
$T_r \leftarrow T_r + (t_r - T_r)/\tau$ **for** $r = 0, \dots, N_N - 1$
$\{v, b\} \leftarrow \{\mathrm{sum}(1/T_{[0:N_N)})/N_N, m_0\}$
**for** $r = 1, \dots, N_N - 1$
$e \leftarrow m_{r-1} + (m_r - b) + [(1/T_{r-1} - v)N/vN_N]$
$\{\mathrm{b}, m_r\} \leftarrow \{m_r, e\}$
**return** $m$

OMP_MPI_DRV($N_N, N, m, r, N_P, \boldsymbol{M}, \tau$)
$t_r \leftarrow$MPI_Wtime()
**for** $p = 0, \dots, N_P - 1$
$\{b_p, e_p\} \leftarrow m_r + [\{p, p+1\}(m_{r+1} - m_r)/N_P]$
**omp parallel for nowait** $i = b_p, \dots, e_p - 1$
$\mathrm{der}_i \leftarrow$Calculate($\boldsymbol{M}_i$)
MPI_Iallgatherv$\left(\mathrm{der}_{[b_p:e_p)}\right)$
$m \leftarrow$MPI_RESCHED4($N_N, N, m, t, \tau$)

### G. Network of Shared Memory Workstations with GPUs

The scheduler for a network of shared memory symmetric multiprocessors with GPUs (OpenMP+MPI+GPU) is described in *Algorithm 5*.

**Algorithm 5** Parallel scheduler OpenMP+GPU+MPI.

INIT_NG($N_N, N_G, N, r$)
$m \leftarrow$INIT_N($N_N, N, m$),
$n, \delta \leftarrow$INIT_G($N_G, m_r, m_{r+1}, \boldsymbol{M}, V, \widehat{\mathbf{C}}, \widehat{\mathbf{D}}$)
**return** $m, n, \delta$

OMP_GPU_MPI_DRV($N_N, N, n, m, r, \boldsymbol{M}, \tau, S, \delta, \delta_{\min}$)
$\{d, b, n_0\} \leftarrow \{n_{N_G+1} - n_0, n_0, m_r\}$
**for** $g = 1, \dots, N_G + 1$
$\{e, b, n_g\} \leftarrow \{n_{g-1} + [(m_{r+1} - m_r)(n_g - b)/d], n_g, e\}$
$t_r \leftarrow$MPI_Wtime()
$\mathrm{der}, n, \delta \leftarrow$OMP_GPU_DRV3($N_G, n, \boldsymbol{M}, S, \delta, \delta_{\min}$)
MPI_Iallgatherv(der)
$m \leftarrow$MPI_RESCHED4($N_N, N, m, t, \tau$)

It combines *Algorithm 3* and *Algorithm 4* with a few differences. The initialization procedure *INIT_NG* evenly distributes oscillators across GPUs on all nodes using the *INIT_G3* and *INIT_N4* procedures described above. The oscillator numbers for each GPU $n$ are recomputed proportionally to the oscillators' range $d$ scheduled to the $r$-th node, and then the procedure *OMP_GPU_DRV3* is called. Only one partition for exchange over the network $(N_P = 1)$ is used because data transfer between the host and GPUs interferes with data transfer between network nodes. Memory used for transfer from GPUs to the host contains valid data only after the synchronization of data transfer is complete. Additional synchronizations necessary for overlapping GPU and network transfers reduce GPU performance. Thus, it was decided to prioritize GPU performance over network performance because the task is more processing than I/O bound. Optimization of this behavior is a subject of future work.

## IV. APPLICATIONS

The described approach has been used for numerous investigations of large nonlinear and non-locally coupled networks described by various realistic and phenomenological models. It has enabled the computation of a large number of dynamical trajectories with different parameters and initial conditions on a small computing cluster and computing grid. Additionally, it has facilitated the analysis of the influence of parameters and initial conditions on the network behavior. More than $10^5$ combinations of parameters and initial conditions were simulated and analyzed in a relatively short time for various networks.

The realistic Terman model [31] for a network of globus pallidus external (GPE) and subthalamic nucleus (STN) neurons was extended from the original tens to 2000-5000 neurons and studied to determine the conditions of pathological synchronization during Parkinson's disease [32]. This model is of type (1) with 6 differential equations and 5 links per neuron. Transition from pathological synchronization to normal incoherent states and vice versa may be achieved by changing the coupling topology and strength. No multi-stability was observed in this model, i.e., no transition from pathological synchronization to normal incoherent states occurred when only initial conditions were changed.

The described approach was extensively used for the investigation of chimera and solitary states in 1D (one-dimensional) networks with approximately 1000 oscillators described by equations (2)-(3). A chimera state is a phenomenon of the coexistence of coherent and incoherent patterns in non-locally coupled networks of identical oscillators [33]. A solitary state is a synchronized state of identical coupled oscillators where some oscillators have different frequencies than the rest. Such a relatively small system provides the possibility to simulate a large number of network instances simultaneously. About 79,000 trajectories with different initial conditions, parameters, link topologies, and coupling strengths [34] were computed and analyzed using the described approach in a computing grid [20]. A diagram illustrating the regions where different chimera states of the 1D Kuramoto model (3) with $m = 0$, $\varepsilon = 1$, and $\mu = 1/2P$ exist in parameter space is shown in Fig. 3a,b. The parameter $\alpha$ represents the phase lag (3), and parameter $r = P/N$, where $P$ is the coupling radius, $N = 1000$ is the total number of oscillators. Oscillators are located on a 1D torus (circle), and each oscillator is coupled with its $P$ neighbors on both sides. Notations h1-h10 represent parameter regions with the number of incoherent network regions equal to 1-10 respectively. Distributions of oscillators' average frequencies in each region at locations A1-A16 are described in plot insets. For a network of 1000 oscillators, the maximum number of incoherent regions found was 16. There are multi-stability regions where networks can transition from a state with one number of incoherent regions to a state with a different number of incoherent regions by changing the initial conditions.

The main difference between two-dimensional (2D) networks and 1D networks is the distribution of network elements in the nodes of a two-dimensional grid. A two-dimensional network of size $N = N_1 \times N_1$ contains $N_1^2$ network elements, where $N_1$ is the number of network elements along a single dimension. The 2D networks require consideration of significantly more network elements compared to the 1D network and have more links and more complex coupling topologies. The application of the proposed approach to 2D networks requires no software modifications, only modified runtime specifications.

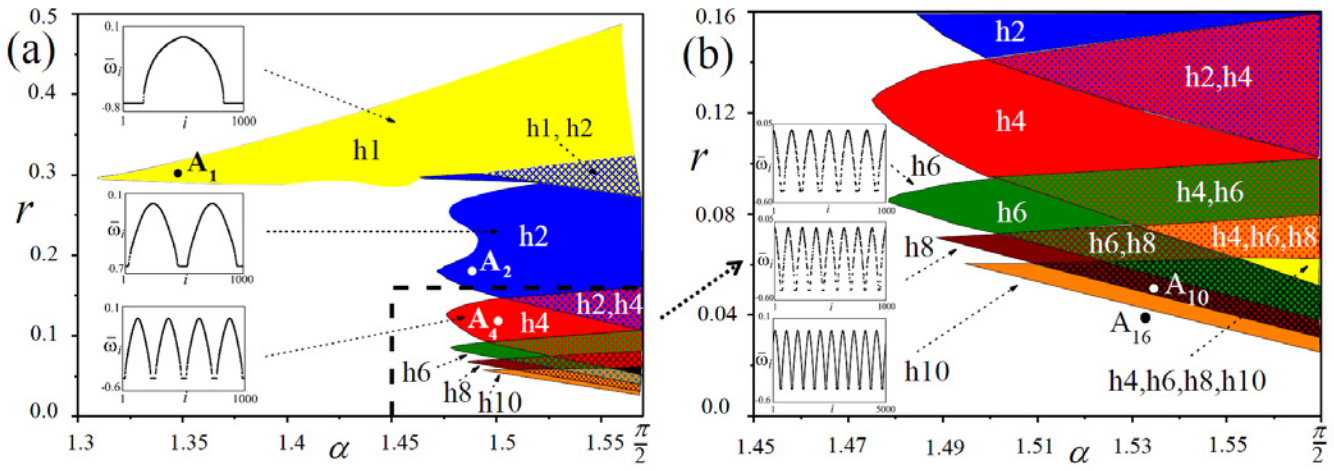

**Fig. 3**. (a) Regions of different chimera states in the parameter space $(r, \alpha)$ for the 1D Kuramoto model (3) [34] with $m = 0$, $\varepsilon = 1$, $\mu = 1/2P$. (b) Enlargement of the rectangle from (a).

The proposed approach was used for the simulation of 2D networks with $10^4$-$10^6$ elements described by models (2)-(3). An example of different chimera and solitary states regions in parameter space $(\alpha, \mu)$ for the 2D Kuramoto model [35] is shown in Fig. 4. The coupling topology used in this study is described by the equation $(i - i')^2 + (j - j')^2 \leq P^2$, i.e., each network element with 2D coordinates $i, j$ is coupled to the network elements with coordinates $i', j'$ located within a circle of radius $P$. The 2D coordinates of network element $i, j$ correspond to the network element numbers in equation (3) as $jN_1 + i$. The distribution of phase in different spiral states is displayed in the insets of Fig. 4. Approximately 27,000 trajectories with different parameters and initial conditions were computed for networks ranging from 100x100 to 800x800 elements in this study.

Elements of three-dimensional networks are distributed in a 3D grid. Each network element has coordinates $i, j, k$, and a network of size $N = N_1 \times N_1 \times N_1$ has $N_1^3$ elements. The 3D coordinates of each network element correspond to the network element number as $kN_1^2 + jN_1 + i$. Simulation of a 3D network requires taking into account more than $10^5$ elements. The described approach was applied to the study of 3D coupled networks described by equation (3) with $10^5$-$10^8$ network elements. An example of chimera state regions for the 3D model (3) with $m = 0$, $\varepsilon = 1$, and $\mu = 1/\Omega$ is displayed in Fig. 5 [36-37], where $\Omega$ is the number of single network element's links. Each network element with coordinates $i, j, k$ is coupled to all network elements with coordinates $i', j', k'$ inside the sphere of radius $P$: $(i - i')^2 + (j - j')^2 + (k - k')^2 \leq P^2$. A corresponding relative radius is $r = P/N_1$. Type I states correspond to oscillating chimeras, i.e., those without spiraling of the coherent region, and Type II are spirally rotating chimeras, called scroll wave chimeras.

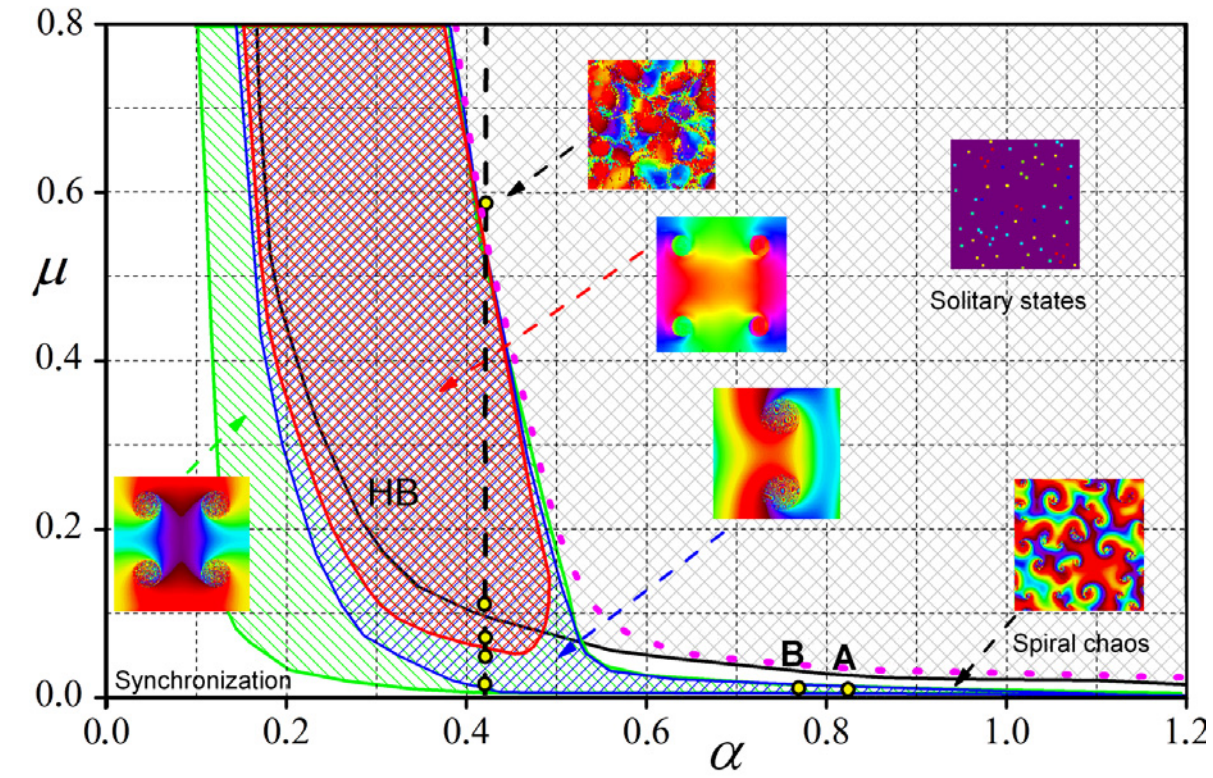

**Fig. 4.** Regions of different spiral chimera and solitary states in the parameter space $(\alpha, \mu)$ for the 2D Kuramoto model (3) with inertia [35].

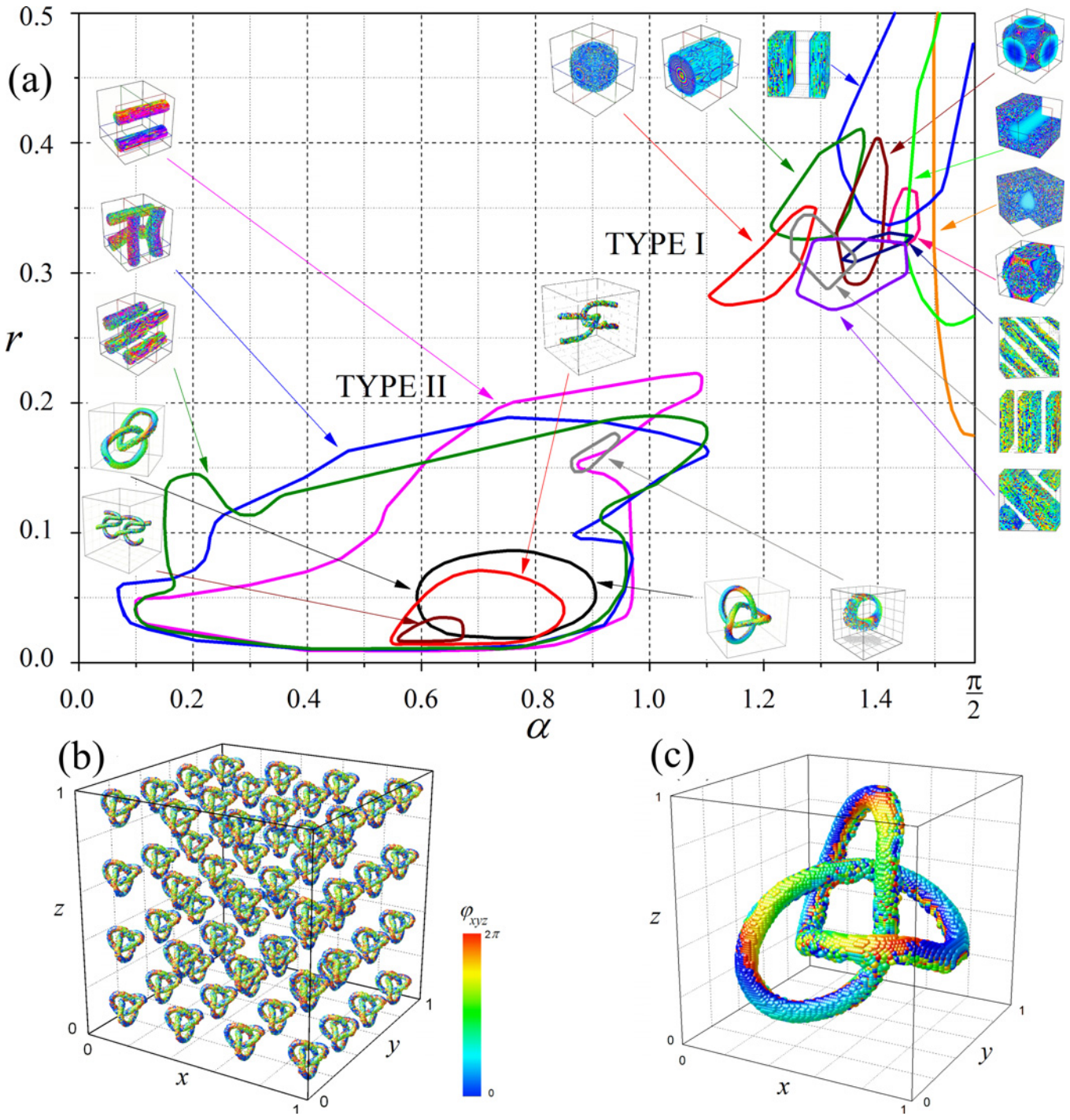

**Fig. 5.** (a) Regions of different chimera states in the parameter space $(r, \alpha)$ for the 3D Kuramoto model (3) [36] with $m = 0$, $\varepsilon = 1$, $\mu = 1/\Omega$, $N_1 = 100$; snapshots of typical chimeras are shown in the insets. (b) Multiple trefoil chimera state [37] $(\alpha = 0.72, r = 0.01, N_1 = 400)$. (c) Trefoil chimera state [37] $(\alpha = 0.68, r = 0.07, N_1 = 100)$.

The number of elements in a 100x100x100 network ($N_1 = 100$) equals one million, and each element has 25-260,000 links. More than 10,000 trajectories were computed in this study for $N_1 = 100$, $N_1 = 200$ and $N_1 = 400$. Phase snapshots of different chimera states are described in the Fig. 5a insets, where the transparent volume corresponds to coherent network elements, while the colored regions correspond to the incoherent ones. An example of a large system with $N_1 = 400$ is presented in Fig. 5b, which was constructed by merging 64 systems with $N_1 = 100$, as shown in Fig. 5c.

The 3D model (3) with inertia ($m \neq 0$) in addition to the solutions presented in Fig. 5 has other chimera and solitary states [38-39] that were discovered using the described approach. Examples of such states are provided in Fig. 6a,b. The scroll toroid chimera state (Fig. 6c) was discovered by the presented approach and is used for performance testing in the next section because this state exists in a relatively wide range of parameters (Fig. 6b), and Kuramoto type models are the most widely used for investigating pattern formation and collective phenomena. This state is also stable to perturbations and can coexist with different chimera and solitary states (Fig. 6d). This property makes the integration step weakly dependent on the model's parameters and mitigates the influence of the integration step on performance.

Estimation of Lyapunov exponents for large networks is a very computationally resource-intensive task that can be solved by the presented approach. The number of Lyapunov exponents characterizes the complexity of the network's chaotic behavior and their computation requires $O(N^2 N_L)$ operations per time step, where $N_L$ is the number of Lyapunov exponents to be estimated, and $N$ is the number of differential equations. Long trajectories ($10^4$-$10^6$ time units) are needed for Lyapunov exponents to converge. The number of positive Lyapunov exponents for the Kuramoto networks is slightly larger than the number of incoherent regions of the networks.

## V. PERFORMANCE

The described parallelizing approach is model-independent to a large degree, but the resulting performance and computing time significantly depend on the number of oscillators and the number of links per oscillator, i.e., problem size, the differential equations, the applied equation solver, hardware, and scheduler. Computing time increases when the differential equation solver selects a short integration step, which is necessary for stiff problems, realistic models, chaotic behavior, etc. Phenomenological models with smooth solutions may be computed faster due to a longer integration step. The number of computing operations per integration step scales as $O(NL)$, where $N$ is the number of network elements, $L$ is the number of links per network element. The number of data transfer operations per integration step scales as $O(N)$. Thus, parallelizing becomes more efficient if the number of links increases.

Reference computing times on different processors and GPUs in shared memory for the models described above are shown in Fig. 7. A confidence interval with a confidence probability of 0.95 is provided. GPUs tend to be faster for larger networks while shared memory multiprocessors offer better performance for smaller networks. Parallelizing is more efficient for larger networks with a higher number of links. For example, computing the realistic Terman model (Section 4) with 2000 oscillators, 6 differential equations, and 5 links per oscillator takes a similar amount of time on 1 and 12 processors as well as on a GPU. The Kuramoto network (3) with $10^6$ oscillators (Fig. 6c) and a coupling radius $P = 2$ has 2 differential equations and 33 links per oscillator. Computing such a network requires a similar amount of time on single and multiple GPUs of the same type, while multiple processors in shared memory provide slightly better performance than a single processor. GPUs are faster than shared memory multiprocessors for this task. For example, an old and slow NVIDIA Quadro K620 GPU outperforms a relatively fast Intel Xeon 8259CL for this run. Newer GPUs are faster than older ones for this task but not significantly. The computing time for networks with a small number of links is mostly determined by the data transfer performance, which leads to worse parallelizing scalability.

The Kuramoto network (3) with $10^6$ oscillators (Fig. 6c) and a coupling radius $P = 6$ has 2 differential equations and 925 links per oscillator. Shared memory multiprocessors and GPUs outperform a single processor for this task. Even multiple GPUs can provide benefits compared to a single GPU and newer GPUs significantly outperform older GPUs. A large number of links significantly increase the percentage of computing operations compared to data transfers and increase parallelizing efficiency.

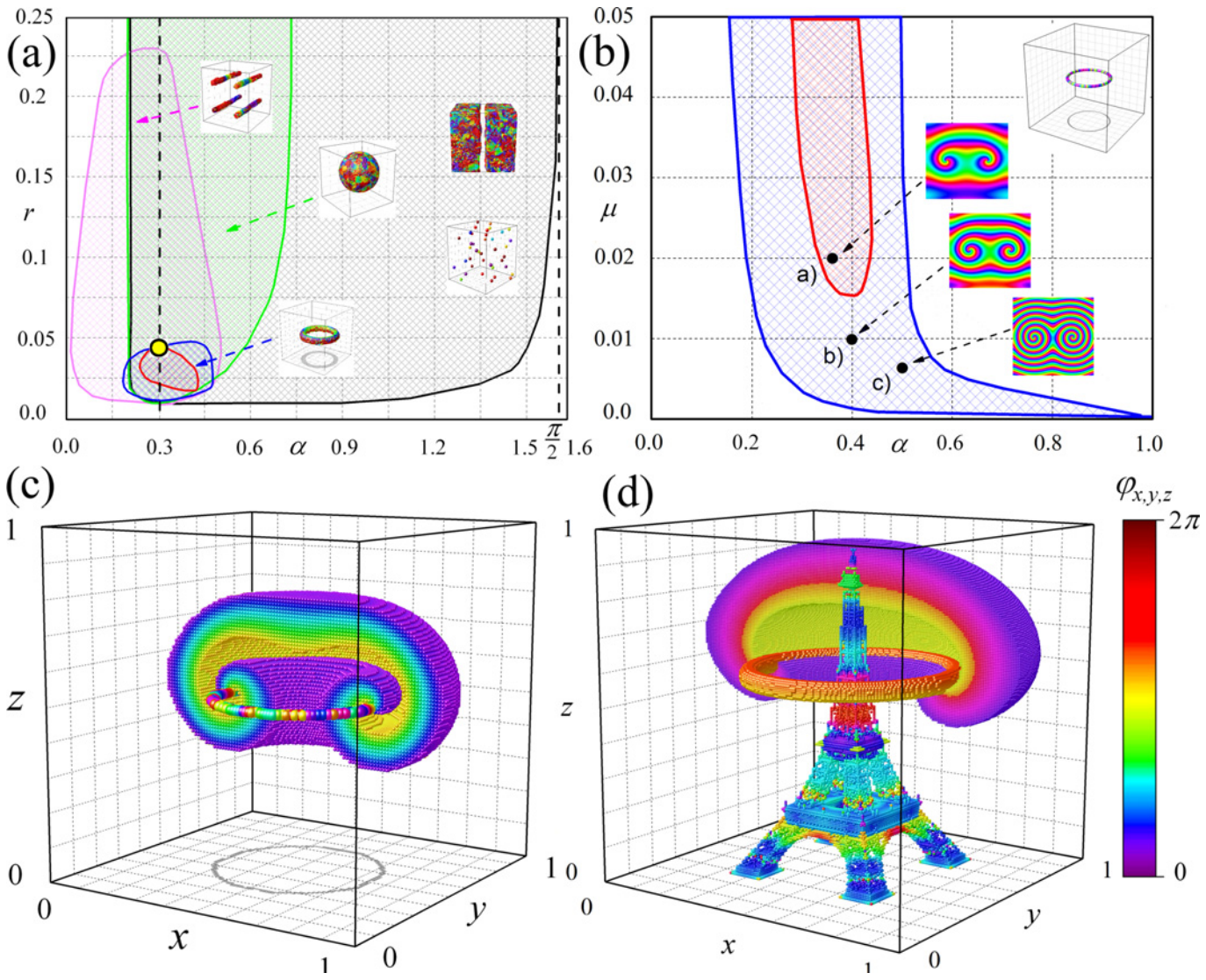

**Fig. 6**. Parameter regions for chimera and solitary states of the Kuramoto model (3) with $m = 1$ and $\varepsilon = 0.05$; typical chimeras are shown in the insets. (a) Parameter plane $(\alpha, r)$ [38], $(\mu = 0.1, N_1 = 100)$. (b) Parameter plane $(\alpha, \mu)$ for scroll ring and toroid chimera states [38], $(r = 0.01, N_1 = 200)$; cross-sections of phases $\varphi_{xyz}$ along $y = 0.5$ are shown in the insets. (c) Scroll toroid chimera state [39] used for performance testing $(\mu = 0.01, r = 0.02, \alpha = 0.45, N_1 = 100)$. (d) Coexistence of solitary states in the form of Eiffel tower with a scroll toroid chimera state [39] $(\alpha = 0.4, r = 0.04, \mu = 0.1, N_1 = 200)$.

Efficiency and speed-up for parallelizing computations of the 3D Kuramoto model with inertia (Fig. 6c) in shared memory are displayed in Fig. 8. Speed-up and efficiency were measured using Intel Xeon 8559CL, 2.5 GHz processors (Fig. 7). The compiler gcc-7.5.0 was used for compilation. Small coupling radii correspond to a small number of links and are the worst cases for parallelizing. For a coupling radius of 2, a single processor is recommended, while for radii 4, 6, and higher, parallelizing efficiency quickly increases and computations may be parallelized for a larger number of processors in shared memory. For $P$=6, the speed-up on 20 processors is about 15 and the efficiency of processor utilization is about 0.75. For large coupling radii and a relatively small number of processors, even super-linear speed-up may take place because processors' cache utilization becomes better.

GPU tests were compiled with NVIDIA CUDA 10.2 libraries using code generation for appropriate computing capabilities. Computation on GPUs is efficient for large networks but using multiple GPUs within the same simulation can only be efficient for very complex problems with a large number of links and network elements. For small tasks the scheduler aims to minimize GPU downtime and schedules computations primarily on the first available GPU to reduce data transfer delay. The computing time and distribution of oscillators per GPU and host for the 3D Kuramoto model (Fig. 6c) with different coupling radii are shown in Fig. 9. The distribution of oscillators corresponds to the last integration step. For a small number of links, computing times on 1, 2, 3, and 4 NVIDIA Turing T4 GPUs are almost the same, as the scheduler maximizes the load on the first GPU. For a larger number of links the utilization of more GPUs provides a reduction in computing time, but not significantly. The host and GPU4 were almost relieved from computations. For the slower NVIDIA Quadro K620 GPU, the host computes about 30% of network elements. Thus for not very large network models, using a single GPU is optimal.

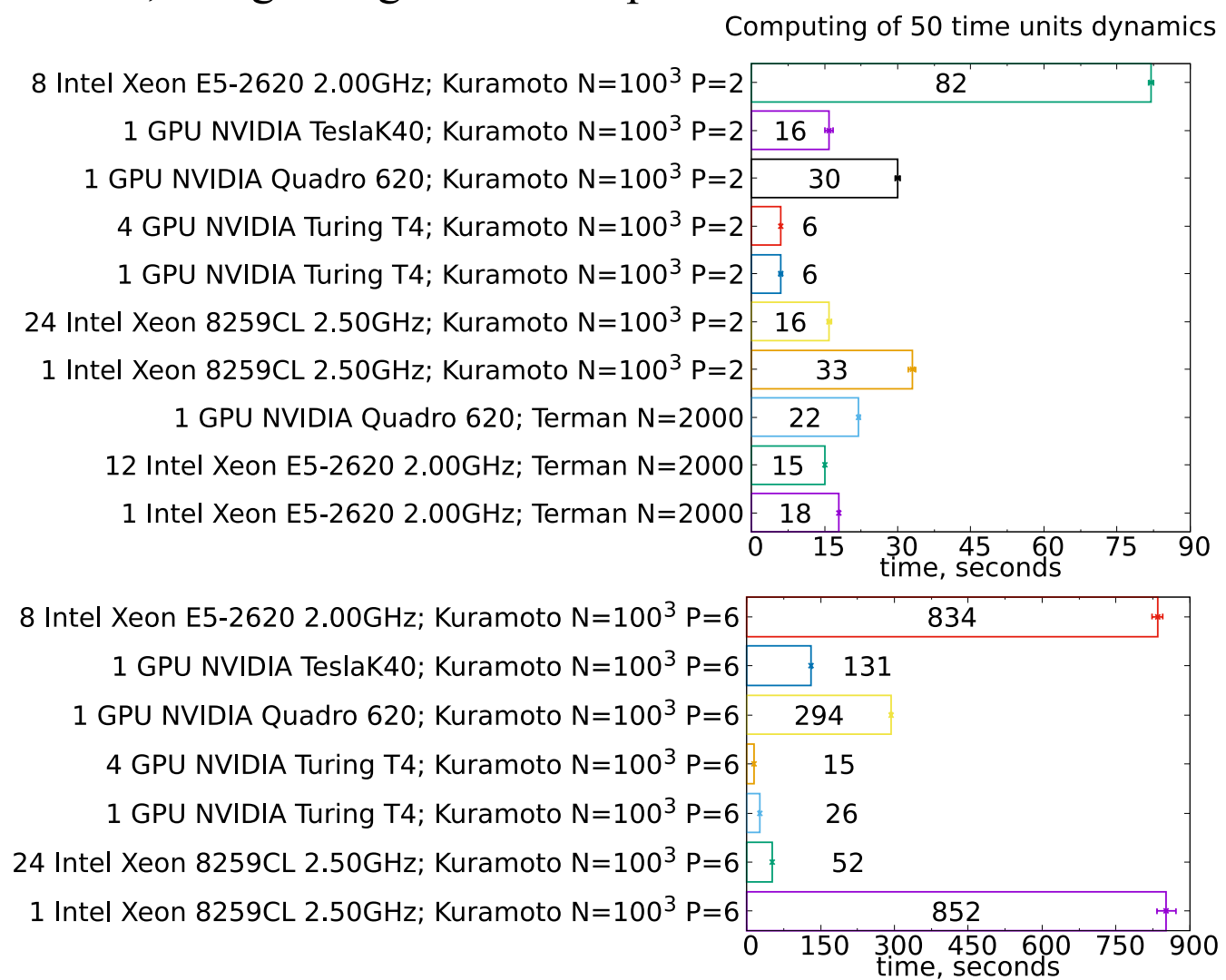

**Fig. 7.** Computation time of 50 time units dynamics for different models on various hardware, with a confidence probability of 0.95.

Parallelizing speed-up and efficiency in distributed memory for the 3D Kuramoto network (Fig. 6c) are displayed in Fig 10. This test was performed using 8 processors (CPUs) per node on up to 8 MPI nodes connected with a Gigabit Ethernet network. OpenMPI-4.1.3 was used as a communication library. Speed-up and efficiency are measured comparatively to the single node with relatively old 8 Intel Xeon E5-2620 2GHz CPUs (Fig. 7). Increasing the number of links significantly increases speed-up and efficiency. Simulations of the network with a small coupling radius cannot be efficiently parallelized for more than 2 MPI nodes in such a computing cluster, while for $P = 6$, up to 7 computing nodes with 56 CPUs may be efficiently used for computations of a single Kuramoto network instance. A speed-up of 40 is achieved in this case, and efficiency is 71%.

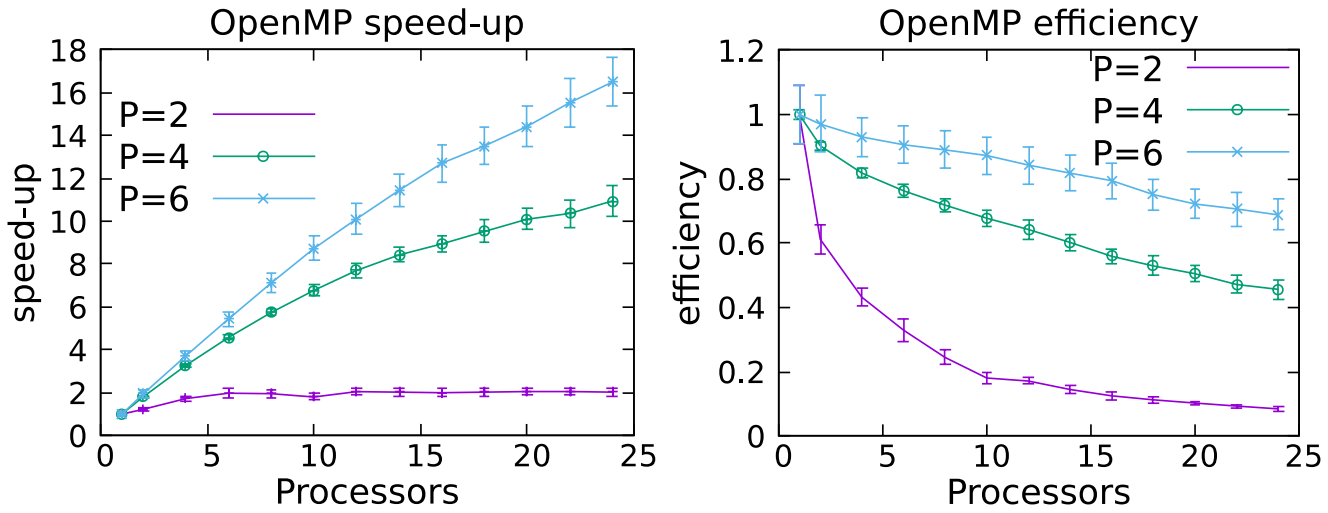

**Fig. 8.** Speed-up and efficiency for the OpenMP scheduler, parallelizing the model (Fig. 6c), with different coupling radii $P$, and confidence probability of 0.95.

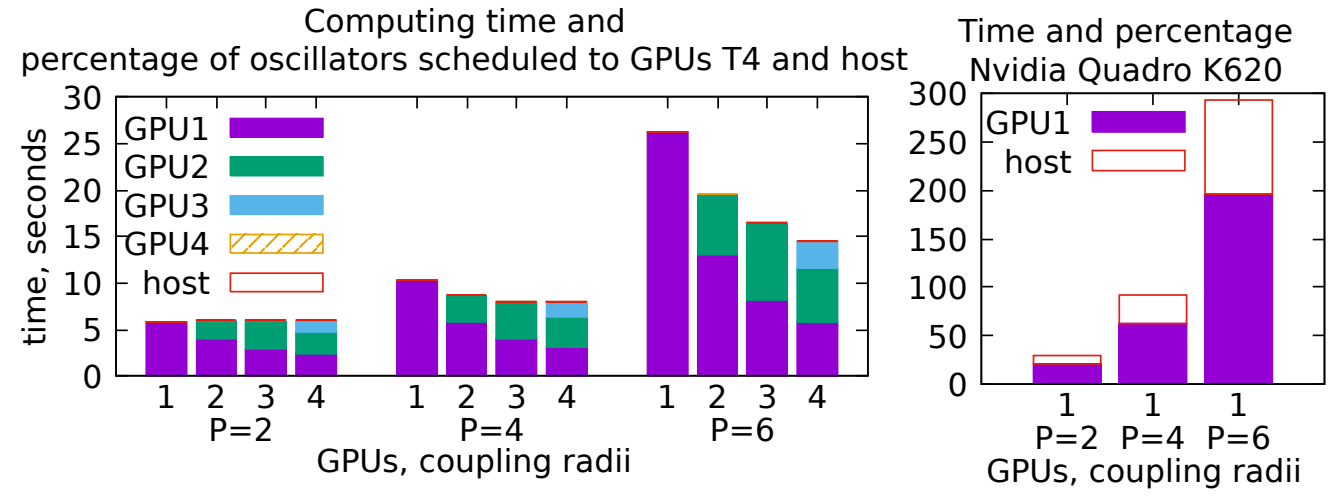

**Fig. 9.** Computing time and scheduling of oscillators per GPU and host for the OpenMP+GPU scheduler, model (Fig. 6c) with 1,000,000 network elements, and different coupling radii $P$, 50 time units dynamics.

Speed-up and efficiency for the utilization of multiple processors and GPUs on different computing cluster nodes for the same run are presented in Fig. 11. The computing cluster nodes had 8 Intel Xeon E5-2620 2GHz CPUs and a single NVIDIA Quadro K620 GPU with 2GB of memory. Speed-up and efficiency were measured relative to a single node (Fig. 7). A slowdown occurs on multiple nodes compared to the single node for a small number of links. For a coupling radius $P = 6$, speed-up may be achieved only for up to 3 cluster nodes. For more nodes, the speed-up is almost the same as on 3 nodes, and efficiency is low. Increasing the number of links increases speed-up and efficiency. As already mentioned in Section 3, the optimization of data exchange over the communication network conflicts with the optimization of data exchange with GPUs and, in the present approach, GPU transfers take precedence. This parallelizing mode may be recommended for large networks of oscillators to compute a single large problem faster. It is also useful when the GPU memory of a single node is not enough for shared memory computations because local and cache variables are distributed across different GPUs of the computing cluster in this mode.

The dependencies of efficiency on the number of processing devices (Fig. 8- Fig. 11) were approximated with the corresponding expression from Amdahl's Law: $1/(\eta c + 1 - \eta)$, where $\eta$ is the percentage of essentially serial operations that cannot be parallelized, and $c$ is the number of processing devices. The aggregated results are provided in Table 1, where $\eta_G$, $\eta_N$, and $\eta_C$ represent the percentage of essentially serial operations related to GPUs, nodes and processors (CPUs); $C_{opt} = 1/\eta_C$, $G_{opt} = 1/\eta_G$, $N_{opt} = 1/\eta_N$ are the optimal number of processors, GPUs, and nodes respectively; $P$ are the coupling radii. The confidence interval for the confidence probability of 0.95 is also specified. The optimal number of processing elements corresponds to efficiency exceeding $1/2$ when parallelizing still makes sense. These cases are highlighted in bold. The speed-up is close to linear if $c \ll 1/\eta$ and $\eta \ll 1$, which are highlighted in gray.

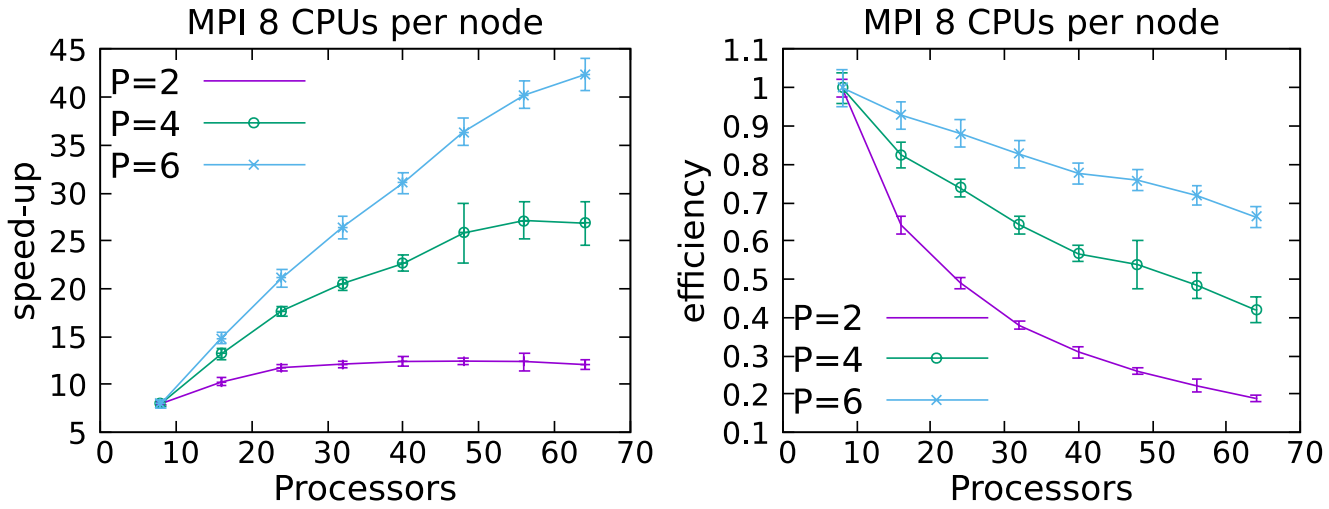

**Fig. 10.** Speed-up and efficiency for the MPI+OpenMP scheduler, 8 processors per node, 1-8 nodes, model (Fig. 6c), different coupling radii $P$, confidence probability of 0.95.

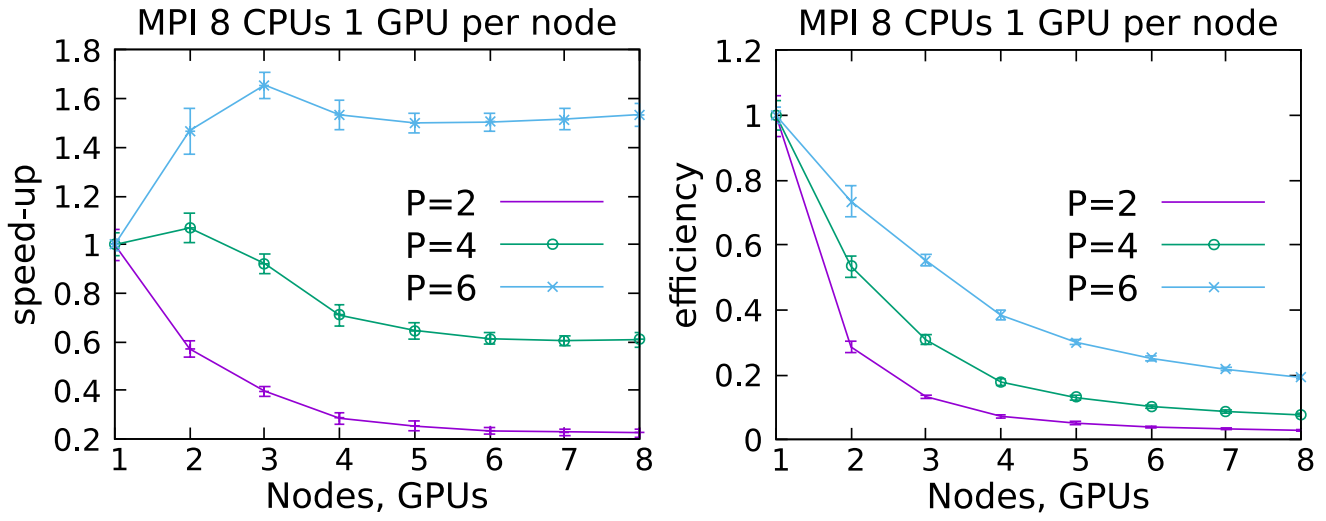

**Fig. 11.** Speed-up and efficiency for the MPI+OpenMP+GPU scheduler, model (Fig. 6c), different coupling radii $P$, confidence probability of 0.95.

TABLE I

PARALLELIZATION EFFICIENCY METRICS

| SCHEDULER | METRIC | $P = 2$ | $P = 4$ | $P = 6$ |
|---|---|---|---|---|
| OPENMP | $\eta_C$ | **0.46** ±0.004 | **0.05** ±0.015 | **0.019** ±0.0029 |
| | $C_{opt}$ | **2.15** ±0.017 | **19** ±0.5 | **50** ±8 |
| OPENMP+ GPU T4 | $\eta_G$ | **1.04** ±0.023 | **0.7** ±0.06 | **0.4** ±0.09 |
| | $G_{opt}$ | **0.96** ±0.021 | **1.4** ±0.12 | **2.4** ±0.6 |
| OPENMP+ MPI | $\eta_N$ | **0.58** ±0.004 | **0.16** ±0.016 | **0.07** ±0.007 |
| | $N_{opt}$ | **1.7** ±0.13 | **6** ±0.6 | **15** ±1.6 |
| | $C_{opt}$ | **14** ±1 | **50** ±5 | **110** ±13 |
| OPENMP+ MPI+ GPU K620 | $\eta_G$,$\eta_N$ | 4.7 ±0.04 | 1.7 ±0.02 | **0.59** ±0.01 |
| | $G_{opt}$, $N_{opt}$ | 0.2 ±0.19 | 0.6 ±0.07 | **1.7** ±0.029 |
| | $C_{opt}$ | 2 ±1.5 | 4.7 ±0.06 | **13.6** ±0.23 |

Computing time for the estimation of Lyapunov exponents for the 3D Kuramoto model (Fig. 6c) on different hardware is presented in Fig. 12. The Lyapunov spectrum estimator can use GPUs and networks of workstations for computing the right-hand side of the differential equations. However, for large networks and a high count of exponents, the computing time is not determined by the differential equations solver but by the matrix manipulation code of the Lyapunov exponent estimator [25]. Parallelizing the latter code has only been implemented for shared memory multiprocessors. Better parallelizing for GPUs and networks of workstations is the focus of future work. The current parallel version offers a significant speed-up, allowing the computation of the 50 largest Lyapunov exponents [39] on a 48-CPU workstation for a 3D network with 1,000,000 elements. This was practically impossible with the serial Lyapunov estimator version [25]. For large networks and coupling radii, the computing time for the same number of Lyapunov exponents should not change significantly, as derivative computations are much faster than the Lyapunov exponents estimator.

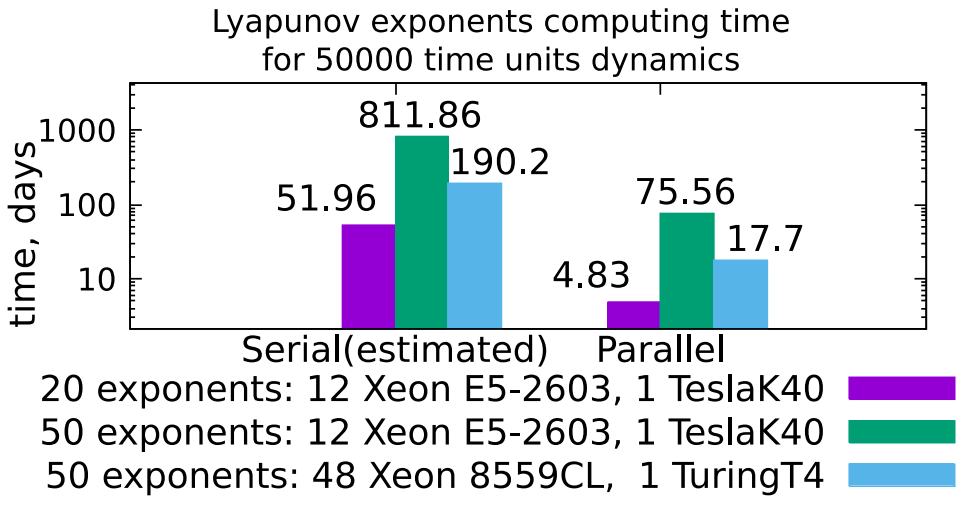

**Fig. 12.** Computing time for Lyapunov spectrum estimation, model (Fig. 6c), coupling radius $P = 2$.

Fig. 13 shows a performance comparison of the presented approach with *BRIAN2*, which is the only neuroscience application we found that supports coupled oscillator networks. Computations were performed on the cluster node used for obtaining data displayed in Fig. 10-Fig. 11. The network (Fig. 6c) has been implemented in *BRIAN2* and simulated using different backends: brian2cpp – a standalone OpenMP C++ application, brian2cithon – a binary single CPU Cython shared object, and brian2cuda – a standalone application for a single GPU-only execution. Backends brian2cpp and brian2cithon of version 2.2 were used to support both fixed and adaptive integration step control. The latest stable version of Brian2cuda-1.0a1 with Brian2-2.4.2 has been used, and they only support fixed integration steps. We found no MPI support for *BRIAN2*; thus, only the OpenMP and OpenMP+CUDA (OpenMP+GPU) versions of the presented approach have been compared.

The left diagram (Fig. 13) displays the execution wall time, i.e., the time from the start until termination of the application for 100 time units of dynamics. This time includes preparation (reading initial data, code generation for *BRIAN2*, construction of the network) and the simulation itself. The right diagram (Fig. 13) displays the simulation time for the last 50 time units of dynamics when all transients are over. The wall time for the presented approach is approximately 2-500 times shorter than that of *BRIAN2*, depending on the backend, parallelization, and integration step control. *BRIAN2* spends most of its wall time on preparation tasks, namely building of the network, and parallelization as well as parameter changes do not noticeably decrease it. The presented approach's wall time decreases as the coupling radius decreases, and parallelization provides a significant speed-up. On a single processor (CPU), the simulation time using *BRIAN2* (Fig. 13 right) is approximately 2-20 times shorter than that of the presented approach. Parallelization does not provide noticeable benefits for *BRIAN2*, while the parallel versions of the presented approach perform simulations as fast as *BRIAN2* for small coupling radii and outperform *BRIAN2* for large coupling radii. Adaptive integration steps, applied in the presented approach, also provide faster execution for *BRIAN2*. Computations on processors with Brian2cpp are about two times faster than with Brian2cithon.

Computation on the GPU using Brian2cuda failed for coupling radii of 4 and 6, with an out-of-memory error occurring after about 10 hours of preparation run. For a coupling radius of 2, Brian2cuda is about 2-10 times slower than the other tested backends. All tests using another GPU backend, Brian2GeNN, failed with an out-of-memory error from the start and are not included in Fig. 13.

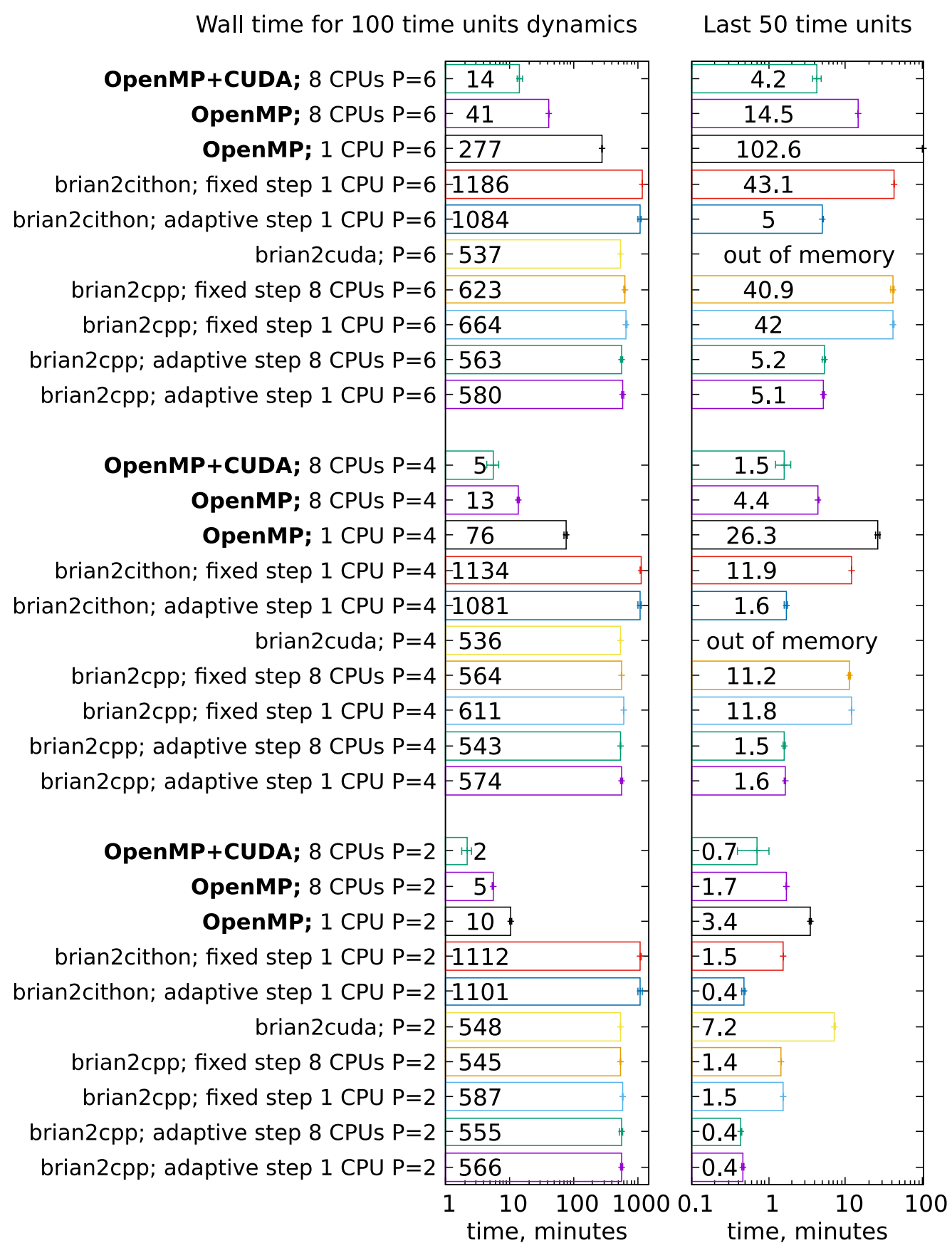

**Fig. 13.** Performance comparison of the presented approach (highlighted in bold) to *BRIAN2* with different backends: wall time for simulating 100 time units of dynamics (left), computing time of the last 50 time units of dynamics (right); model (Fig. 6c); different coupling radii $P$; confidence probability of 0.95.

The dependence of memory resident set size (i.e., not swapped or file-mapped memory) on the coupling radius $P$ (Fig. 14) shows that memory consumption of *BRIAN2* increases as $P^3$, while for the presented approach, the memory consumption is almost independent of the coupling radius in the tested range due to link matrix compression. That is why the Brian2cuda backend failed for $P > 2$ on a GPU with 2 gigabytes of memory, while the presented approach succeeded. Brian2cuda offloads all the computations, including numerical integrator code, to the GPU and transfers the simulation results. The presented approach distributes only the computations of oscillator models across all available GPUs and processors, resulting in more efficient resource usage but introducing additional data transfer overhead. However, the parallelization mitigates this overhead, especially for large coupling radii.

Although we tried to achieve similar test conditions for *BRIAN2* and the presented approach (model, equation solver order, tolerance, integration step duration, etc.), there are differences in the methods used. *BRIAN2* uses a different equation solver and outputs solutions at the integration step boundaries, while our approach interpolates the solution for dense output. Our approach computes the coupled oscillators count (variable *links_amount*, *Algorithm 1*) from the network topology, but we had to hardcode it in the *BRIAN2* implementation. Our approach performs synchronization analysis at each output, which is absent in the *BRIAN2* implementation. Thus, the performance metrics (Fig. 13-Fig. 14 ) demonstrate the trends in performance differences rather than the absolute values.

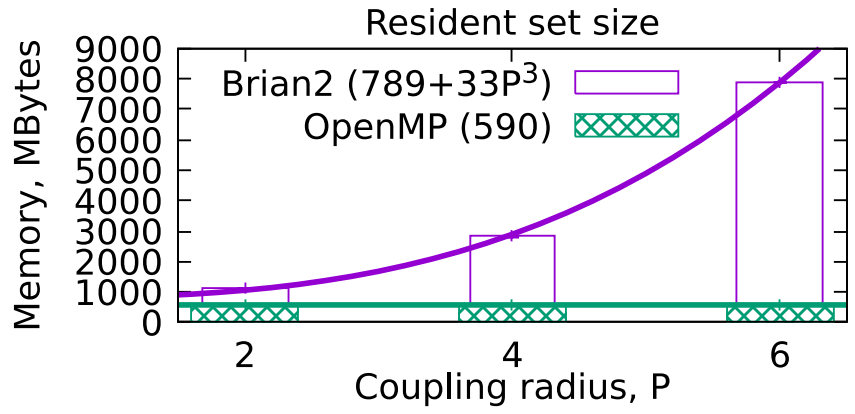

**Fig. 14.** Memory usage comparison of the presented approach (OpenMP scheduler) to *BRIAN2* and fitting its dependence on coupling radius $P$, model (Fig. 6c).

*BRIAN2* generates efficient serial simulation code at the cost of decreased parallelization efficiency, increased preparation time, and higher memory consumption. The latter two become excessively large for large networks with complex topology. Achieving reasonable computing times for large networks with complex topology using *BRIAN2* (Fig. 13) required efforts to examine the generated C++ code and tune Python code. In contrast to *BRIAN2*, the presented approach aims to minimize development and simulation times by enabling easy network model construction and efficient parallelization on different hardware. The runtime overheads introduced by coupling matrix decompression, data transfers, calls of virtual functions, and others are compensated with efficient parallelization that offers performance benefits for large networks of coupled oscillators with complex topologies.

## VI. Discussion and Conclusions

The approach presented in this paper is efficient for parallelizing massive simulations of large nonlinear and non-locally coupled dynamical networks with arbitrary link topologies. Its efficiency is proven by numerous applications for studying nonlinear non-locally coupled dynamical networks with $10^3$-$10^8$ network elements using shared memory multiprocessors, graphics processing units, networks of workstations and combinations of these hardware types. New dynamical network models can be easily integrated without the need for parallel programming by end users.

Although we have mainly focused on investigating fundamental properties of large dynamical networks like chimera and solitary states using Kuramoto and other phase models [34-39], there is no limitation to the application of the described approach to other complex dynamical networks by specifying the oscillator and link models, coupling topology, performing simulations, and aggregating results. Besides phase and phenomenological models, we have successfully applied the proposed approach for the investigation of realistic neuronal network models [32]. We also have plans to use it for realistic modeling of pathological synchronization for epileptic seizures prediction and detection [40], discovering chimera and solitary states in networks of quantum oscillators related to lasers and magnonics, investigation of power grid desynchronization, etc. [41]. The presented approach can be used as a backend for other network simulation software, and the development of specialized APIs, preprocessors, interfaces, and visualization tools, like [18-19], is required for this purpose.

Thanks to efficient resource utilization, the proposed approach may be extremely useful for high-performance network dynamics simulations with modern pluggable edge computing accelerator devices like GPUs in combination with commodity shared memory multiprocessors such as desktops, laptops, microcomputers, etc. Computations may be offloaded to connected edge devices with GPUs like NVIDIA Jetson, or such edge devices may be used as part of a parallel distributed network of workstations with load balancing powered by the presented approach. Due to the dynamic scheduling capacities of the presented approach, online connection and disconnection operations can potentially be implemented for edge devices used for computations. On the other hand, the presented approach can help to efficiently speed up computations on large-scale systems including clusters of workstations, computing grids, clouds, supercomputers, etc.

While GPUs with shared memory and MPI with shared memory implementations achieve significant speed-up, the scalability of current MPI with GPU and shared memory implementation is limited (Table 1) and requires additional optimization, especially for networks with small coupling radii. Such hybrid configurations rely on concurrent accesses of memory buffers by GPUs and network hardware, and these accesses should be synchronized. In the current implementation, the memory buffers are locked (pinned) for concurrent accesses by GPUs, while the network data transfers are delayed until the GPUs' transfers are completed and the buffers contain valid data. An obvious optimization approach to such behavior is locking only part of the buffers for GPUs' data transfers, while transferring previously computed data via the network after synchronization with GPUs. Utilizing the data transfer capabilities of modern GPUs and network adapters like NVLink or GPUDirect RDMA can also be promising. Determining which data transfers could be eliminated may be efficient for optimizing simulations of networks with small coupling radii. Achieving a performance tradeoff with these system-dependent and relatively slow approaches requires additional investigation and is the subject of future work. Nevertheless, the current implementation already allows the use of existing GPU and MPI schedulers without modifications and potentially provides speed-up for large dynamical networks, large coupling radii, small memory, and a small number of network nodes with GPUs. Investigation into the performance optimization of Lyapunov exponent estimation on hybrid hardware will be the subject of future work, as it requires significant modifications to the estimation algorithm [25] that are not related to the dynamical networks.

## VII. Acknowledgment

The authors wish to thank the Ukrainian National Grid infrastructure for providing computing resources. Most computations were performed on the computing cluster of the Information & Computer Centre, National Taras Shevchenko University of Kyiv and the computing cluster of the Technical Center, National Academy of Sciences of Ukraine.

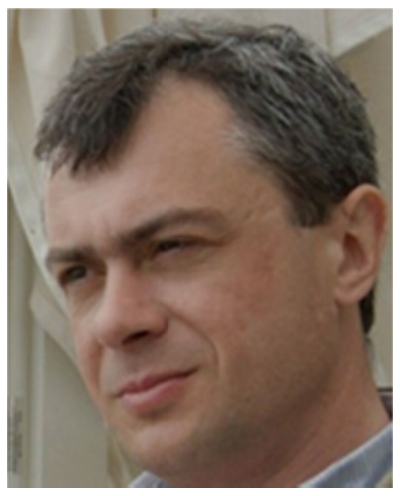

**Oleksandr Sudakov** graduated from the Faculty of Radiophysics at Taras Shevchenko Kyiv University in 1996, where he defended his Ph.D. degree in physical and mathematical sciences in 2002. His Ph.D. thesis focused on the processing of magnetic resonance imaging signals. He is a Senior Researcher at the Laboratory of Mathematical Modeling of Nonlinear Processes at the Technical Center, National Academy of Sciences of Ukraine, and the Head of the Parallel Computing Laboratory at the Information and Computer Center, Taras Shevchenko National University of Kyiv. His research interests include high-performance computing, data analysis, and physical processes in biological systems.

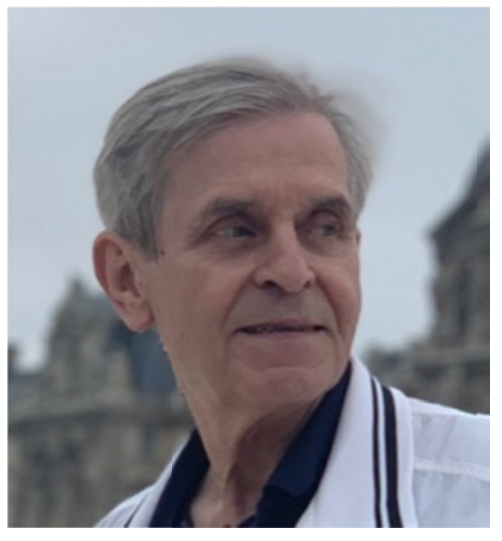

**Volodymyr Maistrenko** graduated from the Faculty of Mechanics and Mathematics at Taras Shevchenko State University of Kyiv in 1978. In 1987, he defended his Ph.D. degree in physical and mathematical sciences at the Institute of Mathematics, National Academy of Sciences of Ukraine. V. Maistrenko worked as a Senior Researcher in the Department of Dynamical Systems at the Institute of Mathematics, National Academy of Sciences of Ukraine. Currently, he is the Head of the Laboratory of Mathematical Modeling of Nonlinear Processes at the Technical Center, National Academy of Sciences of Ukraine. His research interests include Nonlinear Dynamics, Chaos, Networks of Coupled Oscillators, and Chimera states.